\documentclass[onecolumn,showpacs,preprintnumbers,showkeys]{revtex4}
\usepackage{amsthm, amscd, amsfonts, amssymb, graphicx, hhline}
\usepackage{dcolumn}
\usepackage{bm}
\usepackage[english]{babel}
\usepackage{footnpag}
\usepackage{nccfoots}
\usepackage{multirow}
\usepackage[symbol*]{footmisc}

\begin{document}

\title{The role of electron-vibron interaction and local pairing
in conductivity and superconductivity of alkali-doped fullerides.
The route to a room-temperature superconductor
}
\author{Konstantin V. Grigorishin}

\email{konst.phys@gmail.com} \affiliation{Boholyubov Institute for
Theoretical Physics of the National Academy of Sciences of
Ukraine, 14-b Metrolohichna str. Kiev-03680, Ukraine.}
\date{\today}

\begin{abstract}

We investigate the competition between the electron-vibron
interaction (interaction with the Jahn-Teller phonons) and the
Coulomb repulsion in a system with local pairing of electrons on
the triply degenerate lowest unoccupied molecular orbital (LUMO).
The electron-vibron interaction radically changes conductivity and
magnetic properties of alkali-doped fullerides
$\texttt{A}_{n}\texttt{C}_{60}$, which should be antiferromagnetic
Mott insulators: we have found that materials with $n=1,2$ and
$\texttt{A}=\texttt{K},\texttt{Rb}$ are conductors but not
superconductors; $n=3$ and $\texttt{A}=\texttt{K},\texttt{Rb}$ are
conductors (superconductors at low temperatures), but with
$\texttt{A}=\texttt{Cs}$ are Mott insulators at normal pressure;
$n=2,4$ are nonmagnetic Mott insulators. We have shown that
superconductivity, conductivity and insulation of these materials
have common nature. Based on the alkali-doped fullerides we
propose a hypothetical material with a significantly higher
critical temperature using the model of superconductivity with the
external pair potential formulated in a work \emph{K.V.
Grigorishin} Phys. Lett. A \textbf{381} 3089 (2017).
\end{abstract}


 \maketitle

\section{Introduction}\label{intr}

Alkali-doped fullerides ($\texttt{A}_{n}\texttt{C}_{60}$ with
$\texttt{A}=\texttt{K},\texttt{Rb},\texttt{Cs}$ and $n=1\ldots 5$)
demonstrate surprising properties. These materials are
characterized with a narrow bandwidth $W\sim 0.3\ldots
0.5\texttt{eV}$ and a strong one-site Coulomb repulsion $U\sim
0.8\ldots 1.0\texttt{eV}$. Moreover electrons on $t_{1u}$
molecular orbital of fullerene should be distributed according to
Hund's rule: spin of a molecule must be maximal. Thus alkali-doped
fullerides should be antiferromagnetic Mott insulators (the cases
$n=0,6$ a trivial: fulleride $\texttt{C}_{60}$ has empty triply
degeneracy $t_{1u}$ conduction band and alkali-doped fulleride
$\texttt{A}_{6}\texttt{C}_{60}$ has the fully filled conduction
band, hence these materials are band insulators). In reality the
properties of the alkali-doped fullerides are in striking
contradiction to the expected properties. So
$\texttt{A}_{2}\texttt{C}_{60}$ and
$\texttt{A}_{4}\texttt{C}_{60}$ are nonmagnetic insulators. Thus
molecule $\texttt{C}_{60}$ with additional electrons in LUMO does
not have spin, that contradicts to Hund's rule. Under pressure
these materials become metallic. $\texttt{A}_{1}\texttt{C}_{60}$
and $\texttt{A}_{5}\texttt{C}_{60}$ are conductors.
$\texttt{A}_{3}\texttt{C}_{60}$ are superconductors with
$\texttt{A}=\texttt{K},\texttt{Rb}$ for which the critical
temperatures are sufficiently high $T_{c}\sim 30\texttt{K}$.
However for $\texttt{A}=\texttt{Cs}$ the material is insulators,
but it becomes superconductor under high pressure $\sim
2\texttt{kbar}$. The mechanism of superconductivity of these
superconductors has not been fully understood. The positive
correlation between $T_{c}$ and the lattice constant found in
$\texttt{K}$- and $\texttt{Rb}$-doped fullerides has been
understood in terms of the standard BCS theory. Therefore
superconductivity of $\texttt{A}_{3}\texttt{C}_{60}$ is often
described with Eliashberg theory in terms of electron-phonon
coupling and Tolmachev's pseudopotential $\mu^{\ast}$
\cite{gun0,chen1,cap}. However this material shows a surprising
phase diagram \cite{nom}, in which a high transition temperature
of s-wave superconducting state emerges next to a Mott insulating
phase as a function of the lattice spacing.

In the same time there is another approach to describe phases of
alkali-doped fulleride - the model of local pairing
\cite{nom,han1,han2,lam1,lam2,koga}. The experimental basis for
this hypothesis is the fact that the coherence length (size of a
Cooper pair) in the superconducting alkali-doped fulleride is
$\sim 2\ldots 3\texttt{nm}$, which is comparable with a size of a
fullerene molecule $\texttt{C}_{60}$ $\sim 1\texttt{nm}$, that is
a Cooper pair can localize on one molecule and it moves through
lattice by hopping from site to site. Moreover, as noted in
\cite{han1}, the Hubbard-like models predict that
$\texttt{A}_{4}\texttt{C}_{60}$ is an anti-ferromagnetic
insulator, while it is known experimentally that there are no
moments in $\texttt{A}_{4}\texttt{C}_{60}$. The electrons in
$\texttt{A}_{n}\texttt{C}_{60}$ have a relatively strong
interaction with Jahn-Teller intramolecular phonons with $H_{g}$
and $A_{g}$ symmetries. The interaction with the Jahn-Teller
phonons favors a low-spin state and might lead to a nonmagnetic
insulator. At the same time there is, however, a Hund's rule
coupling, which favors a high-spin state. This leads to a
competition between the Jahn-Teller effect and the Hund's rule
coupling. Due to the same cause each molecule
$\texttt{C}_{60}^{-3}$ in $\texttt{Cs}_{3}\texttt{C}_{60}$ crystal
(antiferromagnetic insulator) has spin $S=1/2$ instead $3/2$
\cite{gan}. Proceeding from these facts the local pairing model
suggests that the electron-vibron interaction (interaction with
$H_{g}$ and $A_{g}$ intramolecular Jahn-Teller oscillations)
favors the formation of a local singlet:
$\frac{1}{\sqrt{3}}\sum_{m}C_{im\uparrow}^{+}C_{im\downarrow}^{+}|0\rangle$,
where the spin-up and spin-down electrons is situated on a site
$i$ in the same quantum state $m$. Here $|0\rangle$ is the neutral
$\texttt{C}_{60}$ molecule for the alkali-metal-doped materials,
the quantum number $m$ labels the three orthogonal states of
$t_{1u}$ symmetry. The local singlet state competes with the
normal state (high spin state) of two electrons $
C_{im_{1}\uparrow}^{+}C_{im_{2}\uparrow}^{+}|0\rangle$ dictated by
Hund's rule. Using this assumption it has obtained some important
results. In a work \cite{suz} the density of states in a band
originated from $t_{1u}$ level has been calculated. It has been
found that $\texttt{A}_{2}\texttt{C}_{60}$ and
$\texttt{A}_{4}\texttt{C}_{60}$ are nonmagnetic semiconductors and
the band gaps in these materials are cooperatively formed by the
electron-electron and electron-vibron interactions. On the other
hand, $\texttt{A}_{1}\texttt{C}_{60}$,
$\texttt{A}_{3}\texttt{C}_{60}$ and
$\texttt{A}_{5}\texttt{C}_{60}$ are on the border of the
metal-insulator transition. In a work \cite{han1} considering just
the Hubbard model (the electron-vibron interaction is absent
$g=0$, the one-site Coulomb interaction takes place only), it is
found that the metal-insulator transition in
$\texttt{A}_{3}\texttt{C}_{60}$ takes place at the upper range of
what is believed to be physical values of $U/W$, while for
$\texttt{A}_{4}\texttt{C}_{60}$ this happens at the lower range of
these parameters. This agrees nicely with the fact that
$\texttt{A}_{3}\texttt{C}_{60}$ is a metal but
$\texttt{A}_{4}\texttt{C}_{60}$ is an insulator. However, to
explain why $\texttt{A}_{4}\texttt{C}_{60}$ is not
antiferromagnetic, it needs to include the coupling to the
Jahn-Teller $H_{g}$ phonons. The competition between the
Jahn-Teller effect and the Hund's rule coupling increases $U_{c}$
(critical value of the one-site Coulomb repulsion such that if
$U>U_{c}$, then a material is a Mott insulator) again. The
coupling to the $t_{1u}$ plasmons in
$\texttt{A}_{3}\texttt{C}_{60}$ should lead to an additional
increase of $U_{c}$ in this system. This makes that
$\texttt{A}_{3}\texttt{C}_{60}$ can be a metal.

In a work \cite{han2} it is shown that in
$\texttt{A}_{3}\texttt{C}_{60}$ the local pairing is crucial in
reducing the effects of the Coulomb repulsion and overcoming the
lack of retardation effects. So, for the Jahn-Teller $H_{g}$
phonons the attractive interaction is overwhelmed by the Coulomb
repulsion. Superconductivity remains, however, even for
$U_{vib}\ll U$, and $T_{c}$ drops surprisingly slowly as $U$ is
increased. The reason is as follows. For noninteracting electrons
the hopping tends to distribute the electrons randomly over the
molecular levels. This makes more difficult to add or remove an
electron pair with the same $m$ quantum numbers. However as $U$ is
large $U>W$ the electron hopping is suppressed and the local pair
formation becomes more important. Thus the Coulomb interaction
actually helps local pairing. This leads to new physics in the
strongly correlated low-bandwidth solids, due to the interplay
between the Coulomb and electron-phonon interactions. In such
system the Eliashberg theory breaks down because of the closeness
to a metal-insulator transition. Because of the local pairing, the
Coulomb interaction enters very differently for Jahn-Teller and
non-Jahn-Teller models, and it cannot be easily described by a
Coulomb pseudopotential $g-\mu^{\ast}$. Theoretical phase diagram
for $\texttt{A}_{3}\texttt{C}_{60}$ systems has been obtained with
the E-DMFT analysis in a work \cite{nom}. There are three phase:
the superconducting phase at low temperature, the normal phase at
more high temperatures and the phase of paramagnetic Mott
insulator at bigger volume per $C_{60}^{-3}$. The s-wave
superconductivity is characterized by a local order parameter
$\Delta=\sum_{m=1}^{3}\left\langle
C_{im\uparrow}^{+}C_{im\downarrow}^{+}\right\rangle$, which
describes intraorbital Cooper pairs for the $t_{1u}$ electrons,
$m$ and $i$ are the orbital and site(= molecule) indices
respectively, the site index can be omitted because $\Delta$ does
not depend on a site (the solution is homogenous in space).
Superconducting mechanism is that in such a system we have $U'>U$
($U'$ is interorbital repulsion and $U$ is intraorbital) due the
electron-vibron interaction. Interesting observation is that the
double occupancy $\left\langle
n_{m\uparrow}n_{m\downarrow}\right\rangle$ on each molecule
increases toward the Mott transition and it jumply increases in a
point of transition from the metal phase to the Mott insulator
phase. Conversely, the double occupancy $\left\langle
n_{m\uparrow}n_{m'\downarrow}\right\rangle$ on each molecule and
spin $S$ per molecule decreases toward the Mott transition and
they jumply decreases in the point of transition from the metal
phase to the insulator phase (the spin changes from $3/2$ to
$1/2$). These facts speak in favor of the local pairing
hypothesis. The hypothesis is confirmed with quantum Monte Carlo
simulations of low temperature properties of the two-band Hubbard
model with degenerate orbitals \cite{koga}. It have been clarified
that a superconducting (SC) state can be realized in a repulsively
interacting two-orbital system due to the competition between the
intra- and interorbital Coulomb interactions: it must be $U<U'$.
This s-wave SC state appears along the first-order phase boundary
between the metallic and paired Mott states in the paramagnetic
system. On the other hand, the exchange interaction $K$
destabilizes the SC state additionally.

In the Section \ref{model} based on the local pairing hypothesis
we propose a general approach to description of the properties of
alkali-doped fullerides $\texttt{A}_{n}\texttt{C}_{60}$
($\texttt{A}=\texttt{K},\texttt{Rb},\texttt{Cs}$, $n=1\ldots 5$).
We show that mechanism of superconductivity of
$\texttt{A}_{3}\texttt{C}_{60}$ (interaction with the Jahn-Teller
phonons and the local pairing) uniquely determines conductivity of
$\texttt{A}_{n}\texttt{C}_{60}$ with $n=1,3,5$ and insulation of
the materials with $n=2,4$ with lack of antiferromagnetic
properties. Thus superconductivity, conductivity and insulation of
these materials have common nature. Obtained results can be used
for increase of the critical temperature. In the Section
\ref{temp} based on the alkali-doped fullerides we propose
hypothetical material for realization of model of
superconductivity with the external pair potential formulated in
work \cite{grig}. In this model the energy gap tends to zero
asymptotically as $1/T$ with increasing of temperature. Formally
the critical temperature of such superconductor is equal to
infinity.

\section{The mechanism of conductivity and superconductivity}\label{model}

\begin{figure}[h]
\includegraphics[width=8cm]{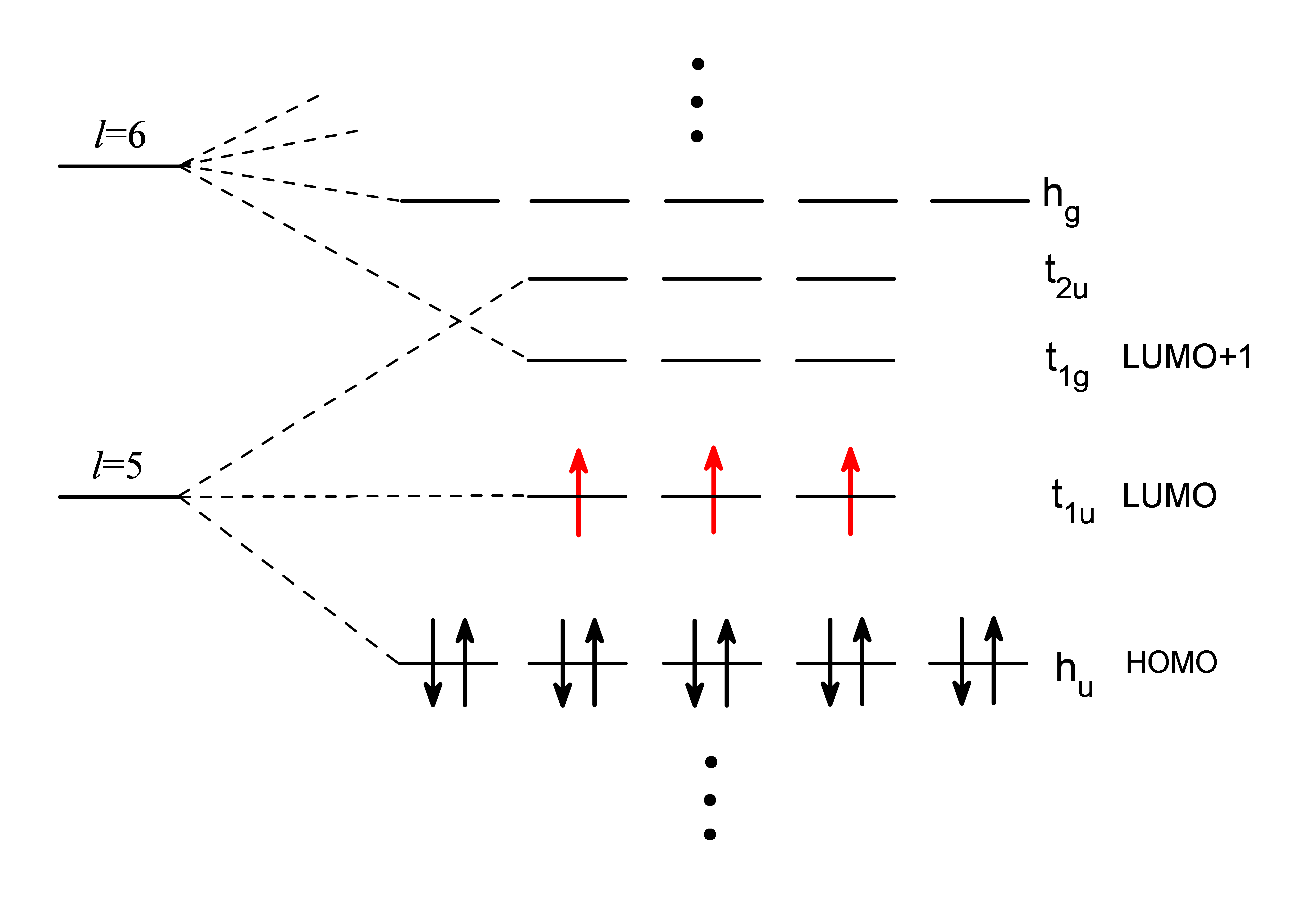}
\caption{The molecular levels of $\texttt{C}_{60}$ close to the
Fermi level. In a substance $\texttt{A}_{3}\texttt{C}_{60}$ the
atoms of alkali metal
$\texttt{A}=\texttt{K},\texttt{Rb},\texttt{Cs}$ give electrons to
the LUMO of the fullerene molecule (red color).} \label{Fig1}
\end{figure}

\begin{figure}[h]
\includegraphics[width=7cm]{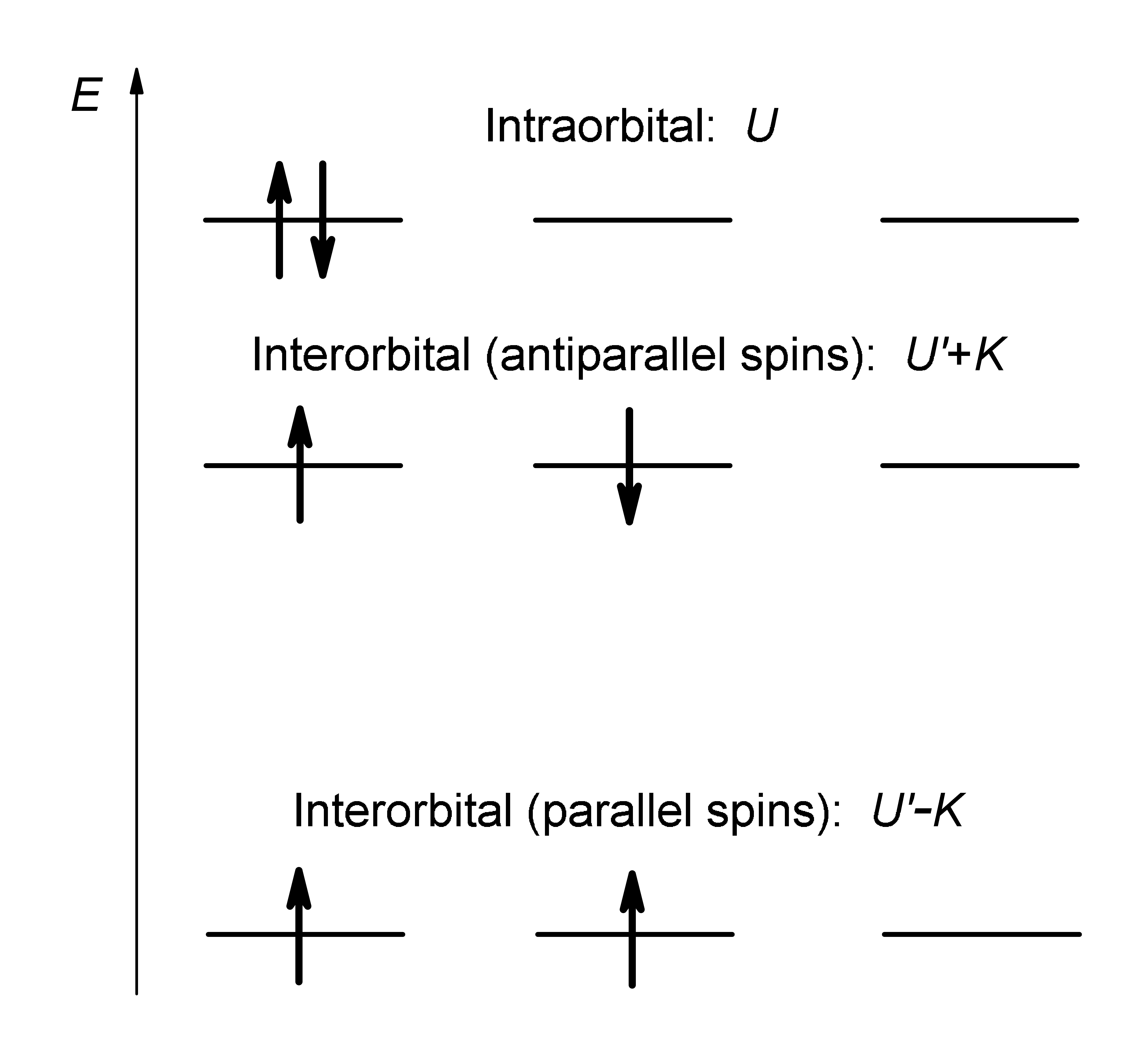}
\caption{The electron configurations of two electrons on LUMO of a
molecule $\texttt{C}_{60}$. Ground state corresponds to
configuration with parallel spins. The intraorbital configuration
has the largest energy.} \label{Fig2}
\end{figure}

Let us consider alkali-doped fulleride
$\texttt{A}_{n}\texttt{C}_{60}$
($\texttt{A}=\texttt{K},\texttt{Rb},\texttt{Cs}$, $n=1\ldots 5$).
Due to the quasispherical structure of the molecule
$\texttt{C}_{60}$ the electron levels would be spherical harmonics
with the angular momentum $l$, however the icosahedral symmetry
generates the splits of the spherical states into icosahedral
representation \cite{sav,allon,hadd}. Fig.\ref{Fig1} shows
molecular levels close to the Fermi level. The LUMO is the 3-fold
degenerate $t_{1u}$-orbital (i.e. it can hold up to six
electrons). It is separated by about 2$\texttt{eV}$ from the
highest occupied molecular orbital (HOMO) and by about
1$\texttt{eV}$ from the next unoccupied level (LUMO+1). The
alkali-metal atoms give electrons to the empty $t_{1u}$ level so
that the level becomes partly occupied. It should be noted that
with spin-orbit coupling taken into account, the static
Jahn-Teller configurations in the $\texttt{C}_{60}^{-}$ molecule
are unstable even in the limit of strong electron-vibron coupling
and the symmetry of the atomic configuration of the unperturbed
$\texttt{C}_{60}$ molecule is restored under time averaging
\cite{rem}. Hamiltonian of the system can be written in a form of
three-orbital Hubbard Hamiltonian with the Hund coupling
\cite{koga,geor} and the electron-vibron (Jahn-Teller phonons)
interaction:
\begin{eqnarray}\label{1.1}
\widehat{H}&=&\sum_{\langle
ij\rangle}\sum_{m}\sum_{\sigma}\left(t_{ijmm}+(\varepsilon_{m}-\mu)\right)
a^{+}_{im\sigma}a_{jm\sigma}+\frac{1}{2}U\sum_{i}\sum_{m}\sum_{\sigma}n_{im\sigma}n_{im-\sigma}\nonumber\\
&+&\frac{1}{2}\left(U'-K\right)\sum_{i}\sum_{m<m'}\sum_{\sigma}n_{im\sigma}n_{im'\sigma}
+\frac{1}{2}\left(U'+K\right)\sum_{i}\sum_{m<m'}\sum_{\sigma}n_{im\sigma}n_{im'-\sigma}\nonumber\\
&+&\frac{1}{2}V\sum_{\langle
ij\rangle}\sum_{mm'}\sum_{\sigma\sigma'}n_{im\sigma}n_{jm'\sigma'}+\widehat{H}_{vib},
\end{eqnarray}
where $a^{+}_{im\sigma}(a_{im\sigma})$ is the electron creation
(destruction) operator in the orbital $m=1,2,3$ localized at the
site $i$; $n_{im\sigma}=a^{+}_{im\sigma}a_{im\sigma}$ is the
particle number operator; $\sigma$ is the spin index; $t_{ijmm}$
is the hopping integral between neighboring sites and the same
orbitals; $\varepsilon_{m}$ is the orbital energy (the energies
can be supposed equal for all $m$); $\mu$ is the chemical
potential; $U$ is the intra-orbital on-site Coulomb repulsion
energy; $U'$ is the inter-orbital on-site Coulomb repulsion
energy; $K>0$ is the on-site exchange interaction energy:
\begin{eqnarray}
  U &=& \int d\textbf{r}d\textbf{r}'|\phi_{m}(\textbf{r})|^{2}V_{c}(\textbf{r},\textbf{r}')|\phi_{m}(\textbf{r}')|^{2} \label{1.1a}\\
  U' &=& \int d\textbf{r}d\textbf{r}'|\phi_{m}(\textbf{r})|^{2}V_{c}(\textbf{r},\textbf{r}')|\phi_{m'}(\textbf{r}')|^{2} \label{1.1b}\\
  K &=& \int d\textbf{r}d\textbf{r}'\phi_{m}^{\ast}(\textbf{r})\phi_{m'}^{\ast}(\textbf{r}')
  V_{c}(\textbf{r},\textbf{r}')\phi_{m}(\textbf{r}')\phi_{m'}(\textbf{r})\label{1.1c},
\end{eqnarray}
where $V_{c}(\textbf{r},\textbf{r}')$ is a operator of on-site
Coulomb interaction. Corresponding electron configurations for two
electrons is shown in Fig.\ref{Fig2}. It should be noted that
$U\gg K$, $U'\simeq U-2K$ (the RPA interaction parameters taken
from \cite{nom2} is $U\sim
0.82\texttt{eV},U'\sim0.76\texttt{eV},K\sim31\texttt{meV}$). Then
$U'-K<U'+K<U$ that means Hund's rule: the electron configuration
with minimal energy has maximal spin. The energy $V$ is the
Coulomb repulsion between neighboring sites ($i\neq j$):
\begin{eqnarray}\label{1.1d}
  V = \int
  d\textbf{r}d\textbf{r}'|\phi_{m}^{i}(\textbf{r})|^{2}V_{c}^{ij}(\textbf{r},\textbf{r}')|\phi_{m'}^{j}(\textbf{r}')|^{2},
\end{eqnarray}
where $V_{c}^{ij}(\textbf{r},\textbf{r}')$ is a operator of
cross-site Coulomb interaction. In the $\texttt{C}_{60}$ molecule
the electrons is coupled strongly to eight $H_{g}$ and two $A_{g}$
intramolecular Jahn-Teller phonons (electron-vibron interaction)
\cite{han1,han2,suz}. The operator of the el.-vib. interaction has
a form:
\begin{eqnarray}\label{1.2}
\widehat{H}_{vib}=\lambda\sum_{i}\sum_{m\leq
m'}\sum_{\sigma}\sum_{\nu}V^{(\nu)}_{mm'}a^{+}_{im\sigma}a_{im'\sigma}\left(b^{+}_{i\nu}+b_{i\nu}\right),
\end{eqnarray}
where $\lambda$ is a coupling constant, the coupling matrices
$V^{(\nu)}_{mm'}$ are determined by icosahedral symmetry
\cite{han1}. For the coupling to $A_{g}$ phonons the matrix is
diagonal with all the diagonal elements equal to unity, for the
coupling to $H_{g}$ phonons the matrix has both diagonal elements
and nondiagonal elements. Vibrational energies for the $A_{g}$ and
$H_{g}$ modes are within the limits $\omega=393\ldots
2271\texttt{K}$ \cite{gun0,koch,beth}. We can see that the
adiabatic parameter in such a system is $\omega/W\lesssim 1$
unlike the parameter in conventional metals where $\omega/W\ll 1$.
Thus Migdal's theorem is violated in alkali-doped fulleride: the
highest-order diagrams has to be counted for electron-phonon
interaction. However for the multi-band Jahn-Teller phonons a
strong reduction of the vertex diagrams occurs \cite{cap2}, that
is Migdal's theorem is valid formally.

\begin{figure}[h]
\includegraphics[width=7cm]{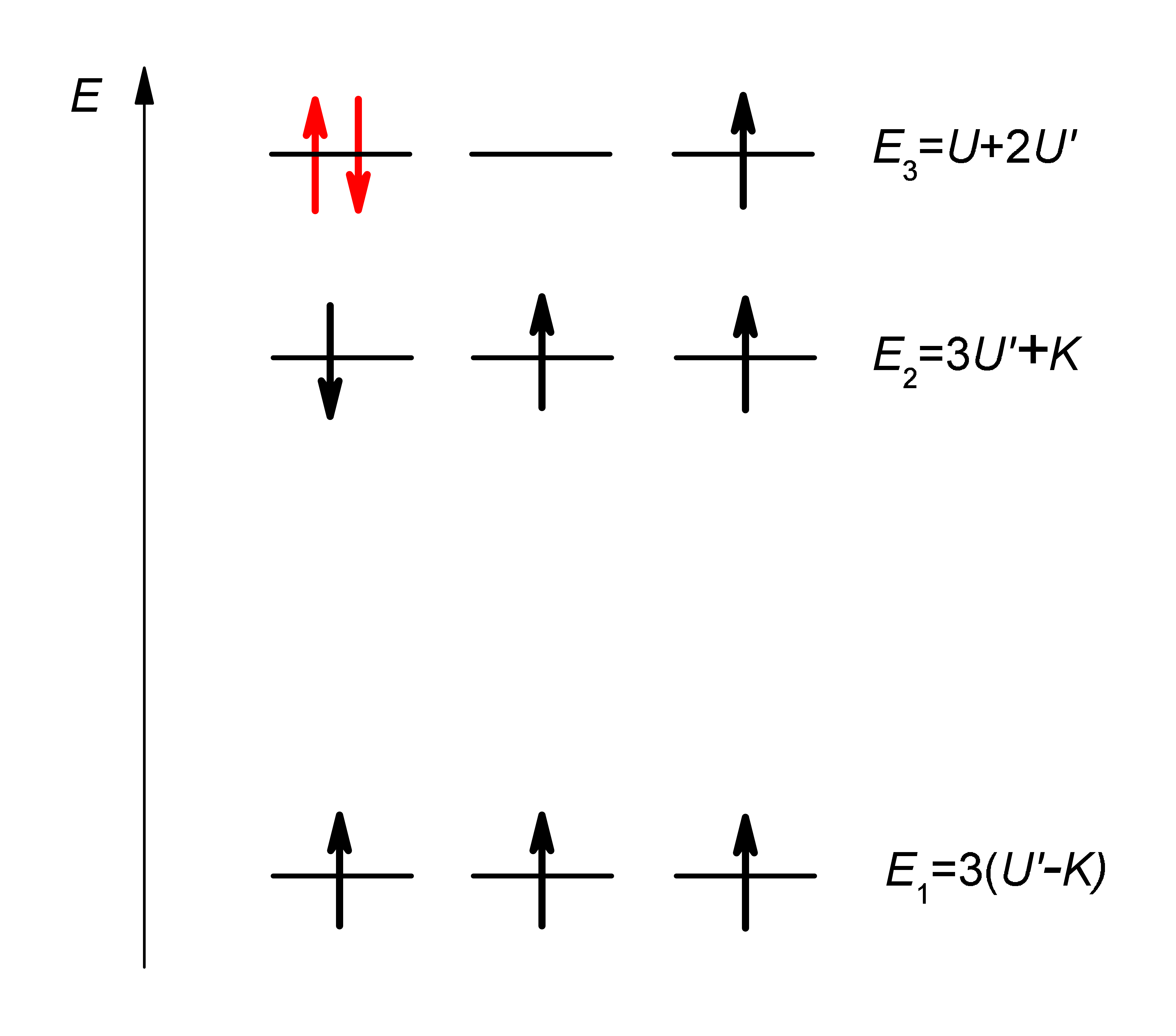}
\caption{The electron configurations of three electrons on LUMO of
a molecule $\texttt{C}_{60}$. Ground state corresponds to
configuration with parallel spins. The configuration with a local
pair (red color) has the largest energy.} \label{Fig3}
\end{figure}

According to the model of superconductivity of
$\texttt{A}_{3}\texttt{C}_{60}$ \cite{han1,han2,lam1,lam2} the
interaction of electrons with $H_{g}$ intramolecular oscillations
favors the formation of a local singlet
\begin{equation}\label{1.4a}
\frac{1}{\sqrt{3}}\sum_{m}C_{m\uparrow}^{+}C_{m\downarrow}^{+}|0\rangle,
\end{equation}
where the spin-up and spin-down electrons have the same $m$
quantum number, i.e., a local pairing takes place. Here
$|0\rangle$ is the neutral $\texttt{C}_{60}$ molecule, the quantum
number $m$ labels the three orthogonal states of $t_{1u}$ symmetry
(LUMO state). In contrast, the normal state (high spin state) of
two electrons is
\begin{equation}\label{1.4b}
\frac{1}{\sqrt{3}}\sum_{m_{1}<m_{2}}C_{m_{1}\uparrow}^{+}C_{m_{2}\uparrow}^{+}|0\rangle.
\end{equation}
As noted in \cite{han2}, for noninteracting electrons the hopping
tends to distribute the electrons randomly over the molecular
levels. However if the on-site Coulomb repulsion $U$ is large so
that $U>W$ the electron hopping is suppressed and the local pair
formation becomes more important. Thus Coulomb interaction
actually helps local pairing and superconductivity is result of
interplay between the Coulomb blockade on a site and the hopping
between neighboring sites. Superconductivity is expected to exist
in this material right up to the Mott transition.

Based on the local pairing approach the Hamiltonian (\ref{1.1})
can be simplified in the following manner. Fig.\ref{Fig3} shows
energies for different electron configurations of
$\texttt{C}_{60}^{-3}$ molecule (that is for the molecule in
$\texttt{A}_{3}\texttt{C}_{60}$ solid). We can see Hund's rule:
the electron configuration with maximal spin $S=3/2$ has minimal
energy: $E_{3}-E_{1}=U-U'+3K\simeq 5K>E_{2}-E_{1}=4K>0$. Thus the
ground state of the system is a state $|m_{1}\sigma, m_{2}\sigma,
m_{3}\sigma\rangle$, in order to form a local pair (\ref{1.4a})
the energy $E_{3}-E_{1}=U-U'+3K$ must be expended. We can measure
energy from the energy of the ground state
$|\uparrow\uparrow\uparrow\rangle$, that is two electrons in the
state $|m\uparrow m\downarrow\rangle$ on a fullerene molecule
"interact" with energy $U-U'+3K$ (the Hund coupling). Then we can
reduce the Hamiltonian (\ref{1.1}) to a form:
\begin{eqnarray}\label{1.3}
\widehat{H}_{eff}&=&\frac{1}{2}\left(U'-K\right)\sum_{i}\langle
n_{i}\rangle\left(\langle n_{i}\rangle-1\right)+
V\sum_{\langle ij\rangle}\langle n_{i}\rangle\langle n_{j}\rangle\nonumber\\
&+&\sum_{\langle
ij\rangle}\sum_{m}\sum_{\sigma}\left(t_{ijmm}+(\varepsilon_{m}-\mu)\right)
a^{+}_{im\sigma}a_{jm\sigma}
+\frac{1}{2}\left(U-U'+3K\right)\sum_{i}\sum_{m}\sum_{\sigma}n_{im\sigma}n_{im-\sigma}+\widehat{H}_{vib},
\end{eqnarray}
where $\langle n_{i}\rangle$ is an average occupation number of a
site $i$. The electron-vibron interaction can be reduce to
BCS-like interaction (point, nonretarded) within single orbital
$m$ that corresponds to diagonal elements of the matrix
$V^{(\nu)}_{mm'}$, and between different orbitals $m$ and $m'$
that corresponds to nondiagonal elements of the matrix. Then the
Hamiltonian takes a form
\begin{eqnarray}\label{1.4}
\widehat{H}_{eff}&=&\sum_{\langle
ij\rangle}\sum_{m}\sum_{\sigma}\left(t_{ijmm}+(\varepsilon_{m}-\mu)\right)
a^{+}_{im\sigma}a_{jm\sigma}\nonumber\\
&+&\left(U-U'+3K-U_{vib}\right)\sum_{i}\sum_{m}n_{im\uparrow}n_{im\downarrow}
-U_{vib}'\sum_{i}\sum_{m'\neq
m}a_{im\uparrow}^{+}a_{im'\uparrow}a_{im\downarrow}^{+}a_{im'\downarrow},
\end{eqnarray}
where we have omitted the ground state energy (the first and
second terms in Eq.(\ref{1.3})). We can distinguish the anomalous
averages $\langle a_{im\downarrow}a_{im\uparrow}\rangle$ and
$\langle a_{im\uparrow}^{+}a_{im\downarrow}^{+}\rangle$ in
Eq.(\ref{1.4}), then the Hamiltonian takes a following form:
\begin{eqnarray}\label{1.5}
\widehat{H}_{eff}&=&\sum_{\langle
ij\rangle}\sum_{mm'}\sum_{\sigma}\left(t_{ij}^{mm}+(\varepsilon_{m}-\mu)\right)
a^{+}_{im\sigma}a_{jm\sigma}\nonumber\\
&+&\sum_{i}\sum_{m}\left[\Delta_{m}^{+}a_{im\uparrow}a_{im\downarrow}
+\Delta_{m}a_{im\downarrow}^{+}a_{im\uparrow}^{+}\right]\nonumber\\
&+&(U_{vib}-U+U'-3K)\sum_{i}\sum_{m}\langle
a_{im\uparrow}^{+}a_{im\downarrow}^{+}\rangle\langle
a_{im\downarrow}a_{im\uparrow}\rangle+U_{vib}'\sum_{i}\sum_{m'\neq
m}\langle a_{im'\uparrow}^{+}a_{im'\downarrow}^{+}\rangle\langle
a_{im'\downarrow}a_{im'\uparrow}\rangle
\end{eqnarray}
where
\begin{eqnarray}\label{1.6}
    \Delta_{m}&=&\frac{U_{vib}-U+U'-3K}{N}\sum_{i}\langle
a_{im\downarrow}a_{im\uparrow}\rangle+\frac{U_{vib}'}{N}\sum_{i}\sum_{m'\neq
m}\langle a_{im'\downarrow}a_{im'\uparrow}\rangle\nonumber\\
\Delta_{m}^{+}&=&\frac{U_{vib}-U+U'-3K}{N}\sum_{i}\langle
a_{im\uparrow}^{+}a_{im\downarrow}^{+}\rangle+\frac{U_{vib}'}{N}\sum_{i}\sum_{m'\neq
m}\langle a_{im'\uparrow}^{+}a_{im'\downarrow}^{+}\rangle
\end{eqnarray}
is the order parameter. $N$ is the number of lattice sites (number
of the molecules). Eq.(\ref{1.6}) means that the order parameter
in such system is determined with the local pairing. We can see
that in this model the exchange energy $3K$ and difference
$U-U'\sim 2K$ resists the electron-vibron interaction
$U_{vib},U_{vib}'$ in the local pairing. As indicated above the
exchange energy $K$ is much less than Coulomb repulsion $U,U'$.
Thus for the local pairing a weaker condition $U_{vib}\gtrsim5K$
must be satisfied than $U_{vib}\gtrsim U,V$. The average number of
electrons per site
\begin{equation}\label{1.7}
    \sum_{m}\langle n_{m}\rangle=\frac{1}{N}\sum_{i}\sum_{m}\sum_{\sigma}\langle
a_{im\sigma}^{+}a_{im\sigma}\rangle
\end{equation}
is determined by the position of the chemical potential $\mu$, and
$\langle\ldots\rangle=Z^{-1}Tr\ldots\exp(-H/T)$ denotes averaging
procedure where $Z$ is a partition function and $T$ is
temperature.

\begin{figure}[h]
\includegraphics[width=16cm]{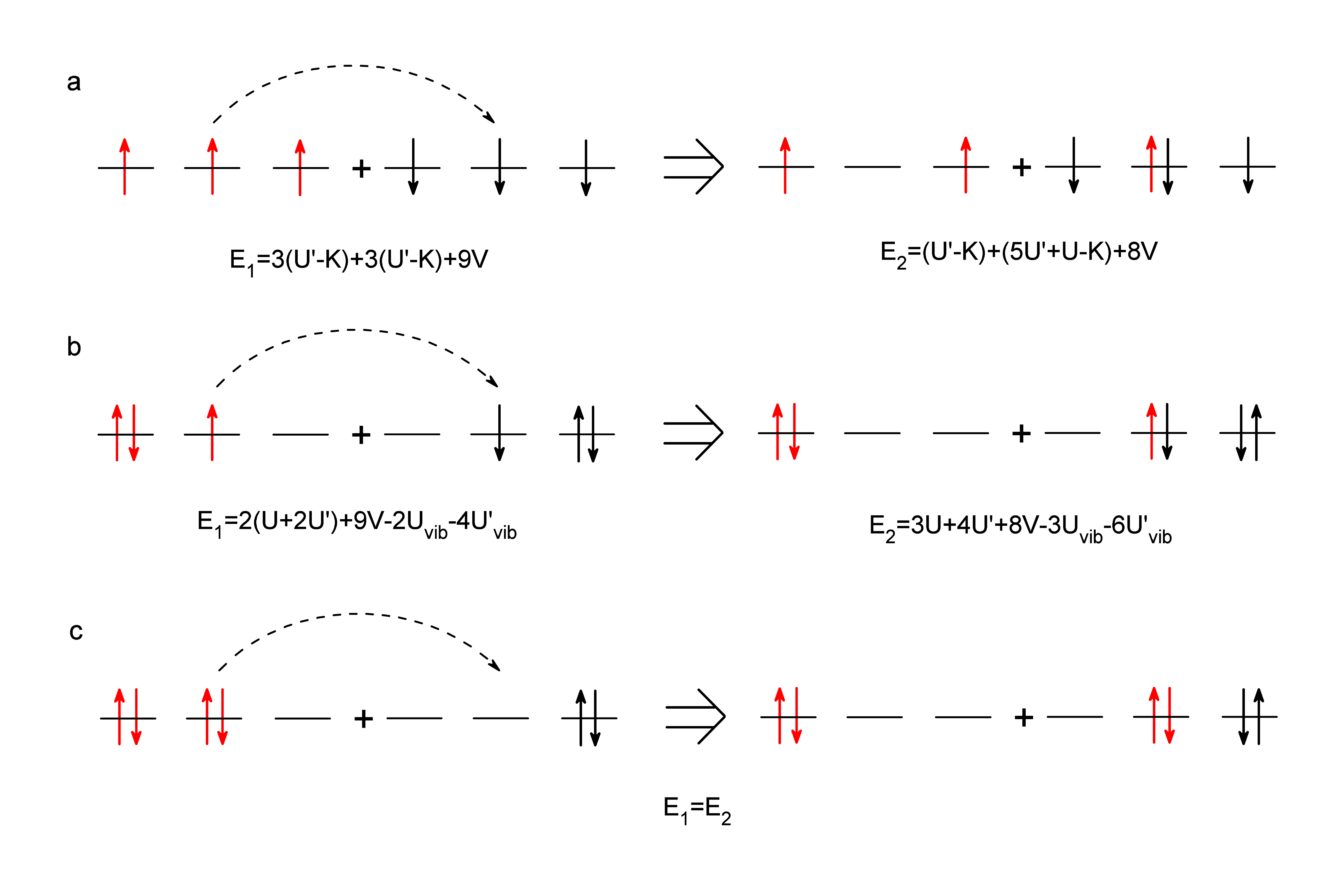}
\caption{The transfer of a charge from a site to a neighboring
site. $E_{1}$ and $E_{2}$ are the energies of the electron
configurations before and after the transfer. (a) - the transfer
of an electron without the el.-vib. interaction, (b) - the same
but with the el.-vib. interaction which forms local pairs on
sites, (c) - the transfer of a local electron pair.} \label{Fig4}
\end{figure}

The on-site Coulomb repulsions $U,U'$, the on-site exchange
interaction energy $K$ and the Coulomb repulsion between
neighboring sites $V$ determine the change of energy at transfer
of an electron from a site to a neighboring site. This process is
shown in Fig.\ref{Fig4}a. We can see that the energy change in
this process is
\begin{equation}\label{1.7a}
    \Delta E=E_{2}-E_{1}=U+4K-V,
\end{equation}
where $E_{1}$ is energy of an initial electron configuration (the
Hund's rule) of neighboring sites, $E_{2}$ is energy of the
configuration if to transfer one electron from the site to
another. A band conductor becomes insulator if
\begin{equation}\label{1.7b}
    \frac{\Delta E}{W}>1,
\end{equation}
where $W$ is a bandwidth. That is an electron does not have enough
reserve of kinetic energy to overcome the Coulomb blockade on a
site. For example let us consider $\texttt{K}_{3}\texttt{C}_{60}$.
According to \cite{nom2}
$W=0.502\texttt{eV},U=0.820\texttt{eV},K=31\texttt{meV},V=0.25\texttt{eV}$,
then $\Delta E/W=2.38$. This means that
$\texttt{K}_{3}\texttt{C}_{60}$ would have to be the Mott
insulator. However $\texttt{K}_{3}\texttt{C}_{60}$ is conductor
and, at low temperatures, is superconductor even.

Accounting of the el.-vib. interaction can change the situation.
For the pair of electrons on a site in state, for example, $m=1$
we have from Eq.(\ref{1.4}):
\begin{eqnarray}\label{1.7c}
\widehat{H}_{vib}(1\leftrightarrow1,2,3)=
&-&U_{vib}a_{1\uparrow}^{+}a_{1\downarrow}^{+}a_{1\downarrow}a_{1\uparrow}
-U_{vib}'\left(a_{1\uparrow}^{+}a_{1\downarrow}^{+}a_{2\downarrow}a_{2\uparrow}
+a_{2\uparrow}^{+}a_{2\downarrow}^{+}a_{1\downarrow}a_{1\uparrow}\right)\nonumber\\
&-&U_{vib}'\left(a_{1\uparrow}^{+}a_{1\downarrow}^{+}a_{3\downarrow}a_{3\uparrow}
+a_{3\uparrow}^{+}a_{3\downarrow}^{+}a_{1\downarrow}a_{1\uparrow}\right).
\end{eqnarray}
Thus each pair obtains energy $-U_{vib}-2U_{vib}'$ due to
interaction within own orbital $1\leftrightarrow 1$ with energy
$-U_{vib}$ and transitions into other two orbitals
$1\leftrightarrow2,1\leftrightarrow 3$ with energy $-U_{vib}'$ for
each. Formation of a local electron pair is possible if electron
configuration with the pair has energy which is less than energy
of electron configuration according to the Hund's rule (see
Fig.\ref{Fig3}): $U-U'+3K-U_{vib}-2U_{vib}'<0$. Then the change of
energy at transfer of an electron from one site to the neighboring
site is
\begin{equation}\label{1.7d}
    \Delta E=E_{2}-E_{1}=U-V-U_{vib}-2U_{vib}',
\end{equation}
as shown in Fig.\ref{Fig4}b. Here we can see more favorable
situation for conductivity because $\Delta E$ in this case is less
than in a case of the Hund's rule configuration. Moreover the
transfer of an electron creates a pair on the neighboring site.
Then, as shown in Fig.\ref{Fig4}c, the pair can move from a site
to a site without change of energy. Thus if the el.-vib. energy is
such that $\Delta E/W<1$, then the material becomes conductor and
superconductor at low temperatures.

\begin{figure}[h]
\includegraphics[width=16cm]{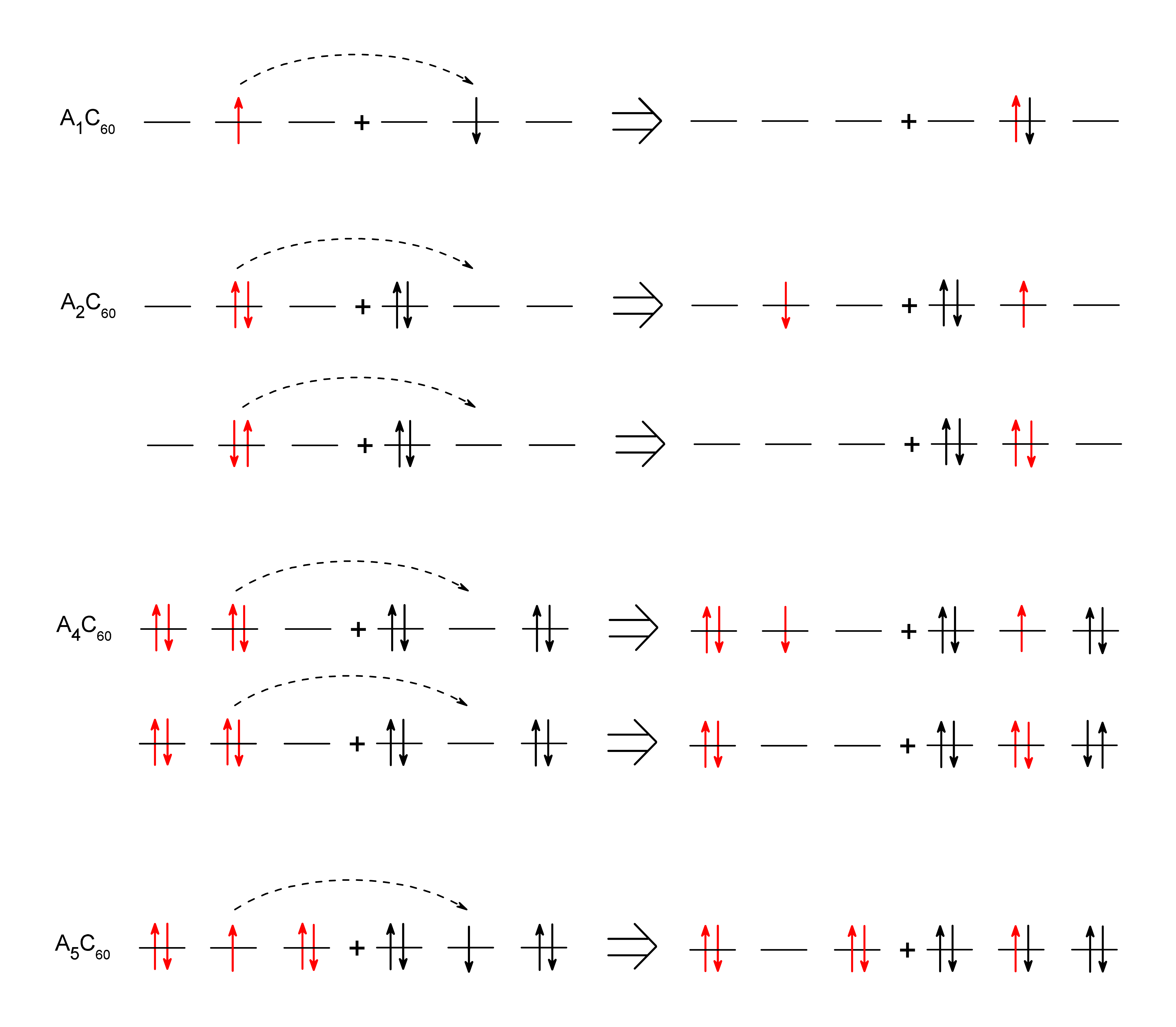}
\caption{The charge transfer in materials
$\texttt{A}_{n}\texttt{C}_{60}$ where $n=1,2,4,5$} \label{Fig5}
\end{figure}

We do not seek the parameters of el.-vib. interaction
$U_{vib},U_{vib}'$ by microscopic calculation but we will
determine them phenomenologically by the following method.
Following a work \cite{litak}, we can make transition from the
site representation (\ref{1.5}) to into the reciprocal (momentum)
space with help of relations:
\begin{equation}\label{1.8}
    a_{\textbf{k}m\sigma}=\frac{1}{\sqrt{N}}\sum_{j}e^{i\textbf{kr}_{j}}a_{jm\sigma},
    \quad
    a_{jm\sigma}=\frac{1}{\sqrt{N}}\sum_{j}e^{-i\textbf{kr}_{j}}a_{\textbf{k}m\sigma},
\end{equation}
then we obtain the effective Hamiltonian like Hamiltonian of a
multi-band superconductor:
\begin{eqnarray}\label{1.9}
\widehat{H}_{eff}&=&\sum_{m}\sum_{\textbf{k}}\sum_{\sigma}\xi_{m}(k)
a^{+}_{\textbf{k}m\sigma}a_{\textbf{k}m\sigma}\nonumber\\
&+&\sum_{\textbf{k}}\sum_{m}\left[\Delta_{m}^{+}a_{\textbf{k}m\uparrow}a_{-\textbf{k}m\downarrow}
+\Delta_{m}a_{-\textbf{k}m\downarrow}^{+}a_{\textbf{k}m\uparrow}^{+}\right],
\end{eqnarray}
with $\xi_{m}(k)=\varepsilon_{m}(k)-\mu$, and two last terms in
Eq.(\ref{1.5}) have been omitted as a constant. The homogeneous
equilibrium superconductivity gaps are defined as
\begin{eqnarray}\label{1.10}
    \Delta_{m}&=&\frac{U_{vib}-U+U'-3K}{N}\sum_{\textbf{k}}\langle
a_{-\textbf{k}m\downarrow}a_{\textbf{k}m\uparrow}\rangle+\frac{U_{vib}'}{N}\sum_{\textbf{k}}\sum_{m'\neq
m}\langle a_{-\textbf{k}m'\downarrow}a_{\textbf{k}m'\uparrow}\rangle\nonumber\\
\Delta_{m}^{+}&=&\frac{U_{vib}-U+U'-3K}{N}\sum_{\textbf{k}}\langle
a_{\textbf{k}m\uparrow}^{+}a_{-\textbf{k}m\downarrow}^{+}\rangle+\frac{U_{vib}'}{N}\sum_{\textbf{k}}\sum_{m'\neq
m}\langle
a_{\textbf{k}m'\uparrow}^{+}a_{-\textbf{k}m'\downarrow}^{+}\rangle,
\end{eqnarray}
and the average number of electrons per site is
\begin{equation}\label{1.11}
    \sum_{m}\langle n_{m}\rangle=\frac{1}{N}\sum_{\textbf{k}}\sum_{m}\sum_{\sigma}\langle
a_{\textbf{k}m\sigma}^{+}a_{\textbf{k}m\sigma}\rangle
\end{equation}
Equations (\ref{1.10}) and (\ref{1.11}) should be solved
self-consistently. It is easy to find that
\begin{equation}\label{1.12}
   \langle
a_{-\textbf{k}m\downarrow}a_{\textbf{k}m\uparrow}\rangle=\frac{\Delta_{m}}{2E_{m}}\tanh\frac{E_{m}}{2T}
\end{equation}
and
\begin{equation}\label{1.13}
    \langle
a_{\textbf{k}m\sigma}^{+}a_{\textbf{k}m\sigma}\rangle=\frac{1}{2}\left(1-\frac{\xi_{m}(k)}{E_{m}}\tanh\frac{E_{m}}{2T}\right),
\end{equation}
where $E_{m}=\sqrt{\xi_{m}(k)+\Delta_{m}^{2}}$. Thus a system with
the local pairing $\langle a_{im\downarrow}a_{im\uparrow}\rangle$
is equivalent to a multi-band superconductor with a continual
pairing $\langle
a_{-\textbf{k}m\downarrow}a_{\textbf{k}m\uparrow}\rangle$ in each
band.

The Eq.(\ref{1.10}) can be simplified using effective
phenomenological parameters. Namely, we can suppose
$U_{vib}=U_{vib}'$. Dispersion low of electrons in $t_{1u}$
conduction band $\xi_{m}(k)$ is supposed the same for different
orbitals $m$. Moreover we suppose the density of states in a
branch $m$ of the conduction band is a constant $\nu=\nu_{0}$ if
$-W/2<\xi<W/2$, otherwise $\nu=0$. Since
$\sum_{\textbf{k}}\rightarrow V\int\nu(\xi)d\xi$ it can be seen
that
\begin{equation}\label{1.14}
    3=\sum_{m}\langle
    n_{m}\rangle=2\frac{V}{N}\sum_{m}\int_{-W/2}^{0}\nu_{0}d\xi\Longrightarrow\nu_{0}=\frac{N}{V}\frac{1}{W}.
\end{equation}
Then Eq.(\ref{1.10}) is reduced to a form
\begin{equation}\label{1.15}
    \Delta=\frac{3U_{vib}-(U-U'+3K)}{W}\int_{-\omega}^{\omega}\frac{\Delta}{2E}\tanh\frac{E}{2T}d\xi.
\end{equation}
Thus we have an effective single-band superconductor with a
coupling constant $g$ determined with el.-vib. interaction
$3U_{vib}$, and the Coulomb pseudopotential $\mu$ determined with
the Hund coupling $U-U'+3K$:
\begin{equation}\label{1.16}
    g=\frac{3U_{vib}}{W}, \quad \mu=\frac{U-U'+3K}{W}.
\end{equation}
The therm $3U_{vib}$ corresponds to the pair's binding energy
$-U_{vib}-2U_{vib}'$ discussed above - Eq.(\ref{1.7c}). The
parameter $U_{vib}$ is chosen so that to obtain the experimental
critical temperatures $T_{c}$ of $\texttt{A}_{3}\texttt{C}_{60}$:
\begin{equation}\label{1.17}
    1=(g-\mu)\int_{-\omega}^{\omega}\frac{1}{2\xi}\tanh\frac{\xi}{2T_{c}}d\xi.
\end{equation}
The rest parameters $W,U,U',K,V$ is taken from \cite{nom2}. The
bandwidth of alkali-doped fulleride is $W\sim 0.5\texttt{eV}$, the
energy gap is $\Delta\approx 2T_{c}\sim30\ldots 60\texttt{K}\ll
\varepsilon_{F}=W/2$ that is a change of the chemical potential at
transition to SC state can be neglected unlike the systems with
the strong attraction and low particle density, where it can be
$\Delta>\varepsilon_{F}$ and the change of the chemical potential
plays important role in formation of SC state \cite{lok}. On the
other hand the vibrational energies for the $A_{g}$ and $H_{g}$
modes are $\omega\sim 0.1eV\sim \varepsilon_{F}$ that means
Tolmachev's weakening of the Coulomb pseudopotential by a factor
$\ln\frac{\varepsilon_{F}}{\omega}$ does not take place. 

The critical temperature $T_{c}$ is calculated for the limit
values of vibration energies of the intermolecular modes $H_{g}$
and $A_{g}$. In addition we calculate the Mott parameter
(\ref{1.7b}) for the state with local pairs (\ref{1.4a}) and for
state without el-.vib. interaction where an electron configuration
corresponds to the Hund's rule (\ref{1.4b}). Results of the
calculations for $\texttt{A}_{3}\texttt{C}_{60}$ (where
$\texttt{A}=\texttt{K},\texttt{Rb},\texttt{Cs}$ and the substance
with cesium is considered at normal pressure, $2\texttt{kbar}$ and
$7\texttt{kbar}$ with a lattice constant $a$ and $\texttt{fcc}$
structure) are presented in the following table:
\begin{center}\label{tab1}
\begin{tabular}{|c|c|c|c|c|c|}
  \hline
    & K & Rb & Cs(7\texttt{kbar}) & Cs(2\texttt{kbar}) & Cs \\
  \hline
  $a(\texttt{A})$ & 14.240 & 14.420 & 14.500 & 14.640 & 14.762 \\
  \hline
  $W(\texttt{eV})$ & 0.502 & 0.454 & 0.427 & 0.379 & 0.341 \\
  \hline
  $U(\texttt{eV})$ & 0.82 & 0.92 & 0.94 & 1.02 & 1.08 \\
  \hline
  $U'(\texttt{eV})$ & 0.76 & 0.85 & 0.87 & 0.94 & 1.00 \\
  \hline
  $V(\texttt{eV})$ & 0.25 & 0.27 & 0.28 & 0.29 & 0.30 \\
  \hline
  $K(\texttt{meV})$ & 31 & 34 & 35 & 35 & 36 \\
  \hline\rule{0cm}{0.45cm}
  $\frac{\Delta E}{W}=\frac{U+4K-V}{W}$ & 1.38 & 1.73 & 1.87 & 2.30 & 2.71 \\[0.1cm]
  \hline
  $T(\texttt{K})$ & 19 & 29 & 35 & 26 & - \\
  \hline
  \rule{0cm}{0.4cm}
  \multirow{2}{*}{$U_{vib}(\texttt{eV})$}$\qquad\omega=393\texttt{K}$ & 0.104 & 0.113 & 0.114 & 0.106 & \multirow{2}{*}{$\sim 0.1$} \\
  $\qquad\qquad\qquad\omega=2271\texttt{K}$ & 0.085 & 0.091 & 0.092 & 0.089 &  \\[0.1cm]
  \hline
  \rule{0cm}{0.5cm}
  \multirow{2}{*}{$\frac{\Delta E}{W}=\frac{U-V-3U_{vib}}{W}$}$\quad\omega=393\texttt{K}$ & 0.51 & 0.69 & 0.74 & 1.09 & \multirow{2}{*}{$\sim 1.4$} \\
  $\quad\qquad\qquad\qquad\qquad \omega=2271\texttt{K}$ & 0.63 & 0.83 & 0.90 & 1.22 &  \\[0.1cm]
  \hline
\end{tabular}
\end{center}
We can see that without the el.-vib. interaction we have
$\frac{\Delta E}{W}>1$, hence all materials are Mott insulators.
The el.-vib. interaction changes the relation as $\frac{\Delta
E}{W}<1$ for $\texttt{A=K,Rb,Cs(at 7kbar)}$, hence these materials
becomes conductors. We can see that the el.-vib. interaction
$U_{vib}$ is approximately constant $\sim 0.1\texttt{eV}$ (that
corresponds to $g\sim 0.7$) at different lattice constants $a$,
and its value is such that $U_{vib}>U-U',K$ but $U_{vib}<U,U'$. A
material $\texttt{Cs}_{3}\texttt{C}_{60}$ at pressure
$2\texttt{kbar}$ is very close to Mott transition, hence our
mean-field approximation is not correct due very strong
fluctuations. For $\texttt{Cs}_{3}\texttt{C}_{60}$ at normal
pressure the el.-vib. interaction cannot ensure the condition
$\Delta E/W<1$, hence this material is a Mott insulator.

For understanding of properties of materials
$\texttt{A}_{n}\texttt{C}_{60}$, where $n=1,2,4,5$, we can suppose
that the el.-vib. interaction $U_{vib}$ is approximately constant
for these materials and is equal for the value in
$\texttt{A}_{3}\texttt{C}_{60}$ $\sim 0.1\texttt{eV}$, it is
analogously for Coulomb $U,U',V$ and exchange interactions $K$.
The charge transfer in these materials is shown in Fig.\ref{Fig5}.

\begin{itemize}
    \item $\texttt{A}_{1}\texttt{C}_{60}$. In order to form a pair we have to transfer an electron from a site
    to a neighboring site containing another electron. For this it is
    necessary to make a positive work $U-V-3U_{vib}>0$. Thus formation of the pairs is energetically unfavorable.  In the same
    time $\frac{\Delta
    E}{W}=\frac{U-V-3U_{vib}}{W}<1$, however $\frac{U-V}{W}\simeq \frac{U'-K-V}{W}\simeq \frac{U'+K-V}{W}>1$ without the el.-vib. interaction.
    Thus this material is conductor due the el.-vib. interaction but it is not superconductor.
    \item $\texttt{A}_{2}\texttt{C}_{60}$. Two electrons are in the paired state on a site because the energy of a paired state is $U-U'-K-3U_{vib}<0$.
    We can transfer the charge from site to site by two ways. (a)
    - by transferring one electron with breaking of a pair. In this
    case $\frac{\Delta E}{W}=\frac{2U'-U-V+3U_{vib}}{W}>1$ too. (b) - by transferring the
    pair. In this case $\frac{\Delta E}{W}=\frac{2U'-4V}{W}>1$. Thus this material is
    insulator and a molecule $\texttt{C}_{60}^{-2}$ does not have spin.
    \item $\texttt{A}_{4}\texttt{C}_{60}$. Like the previous material four electrons are in the paired state on a site
    because the energy of a paired state is negative. We can transfer the charge from site to site by transferring one electron with breaking of
    pair and by transferring the pair. In these cases $\frac{\Delta E}{W}$ is $\frac{2U'-U-V+3U_{vib}}{W}>1$ and $\frac{2U'-4V}{W}>1$ accordingly.
    Thus this material is insulator and a molecule $\texttt{C}_{60}^{-4}$ does not have spin.
    \item $\texttt{A}_{5}\texttt{C}_{60}$. In order to transfer an electron from a site
    to a site we have to expend such energy that $\frac{\Delta E}{W}=\frac{3U-2U'-V-3U_{vib}}{W}<1$.
    This process forms the pair which can move through the system, but $3U-2U'-V-3U_{vib}>0$ (energy of the pair is positive) that is the Cooper pairs are not stable.
    Thus this material is conductor due to the el.-vib. interaction but it is not superconductor.
\end{itemize}

We can see the el.-vib. interaction transforms Mott insulators
$\texttt{A}_{n}\texttt{C}_{60}$ with $n=1,3,5$ to conductors.
However for $n=2,4$ this interaction prevents conduction and it
pairs electrons within each molecules so that the molecules do not
have spin.

\section{The model of a hypothetical room-temperature superconductor.}\label{temp}

\begin{figure}[h]
\includegraphics[width=8cm]{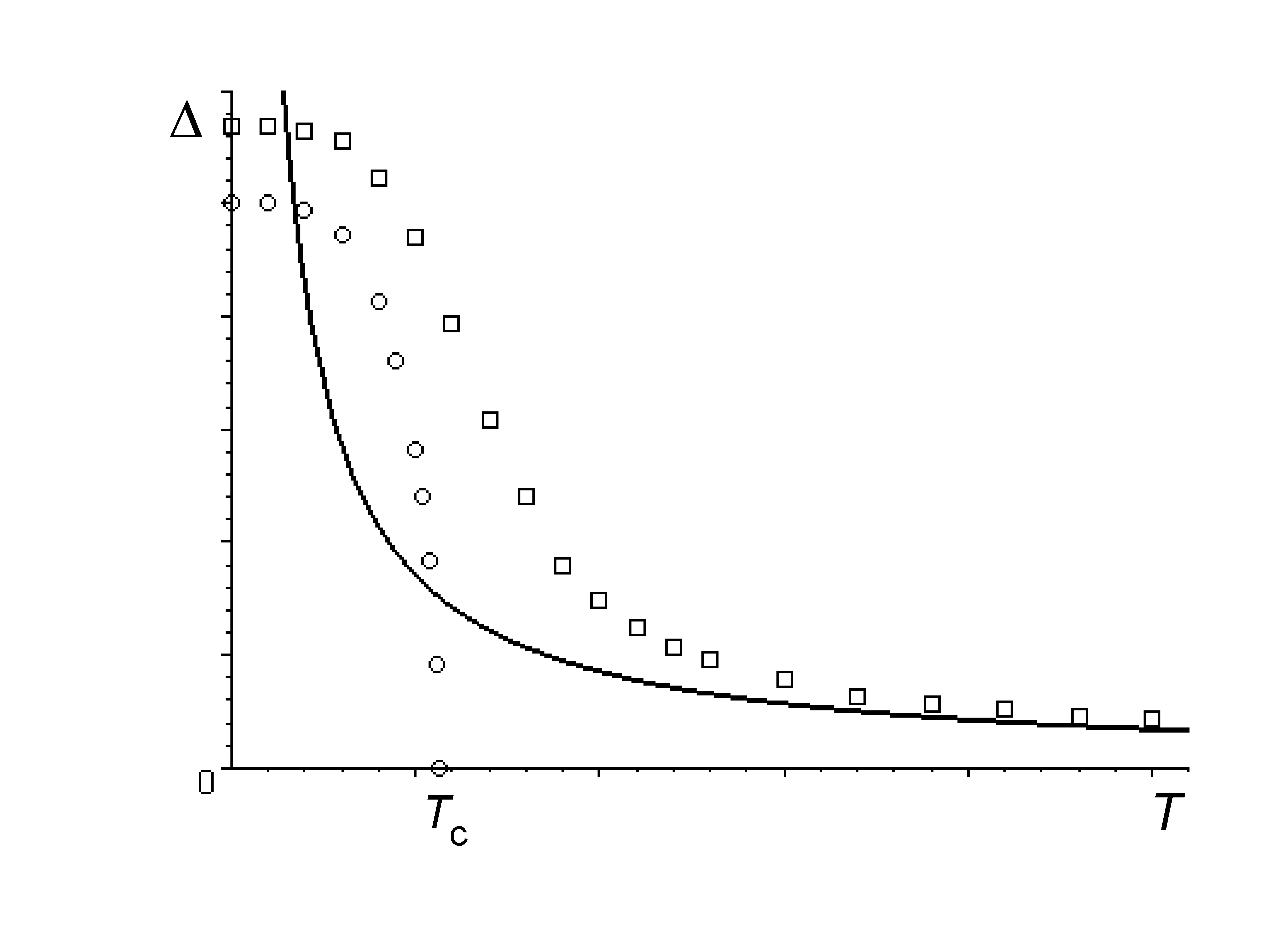}
\caption{the energy gap as function of temperature. Circles are
solutions of Eq.(\ref{2.2}) if $\upsilon=0$ - standard result of
BCS theory with the second order phase transition. Squares are
solutions of Eq.(\ref{2.2}) if $\upsilon>0$. The solutions
demonstrate, that the energy gap tends to zero asymptotically as
temperature rises. Bold line is asymptotical solution (\ref{2.3})
at large temperature.} \label{Fig6}
\end{figure}

In a work \cite{grig} the model of a hypothetical superconductor
has been proposed. In this model an interaction energy between
(within) structural elements of condensed matter (molecules,
nanoparticles, clusters, layers, wires etc.) depends on state of
Cooper pairs: if the pair is broken, then energy of the molecular
system is changed by quantity $\upsilon=E_{a}-E_{b}$, where
$E_{a}$ and $E_{b}$ are energies of the system after- and before
breaking of the pair accordingly. Thus to break the Cooper pair we
must make the work against the effective electron-electron
attraction and must change the energy of the structural elements:
\begin{equation}\label{2.1}
    2|\Delta|\longrightarrow 2|\Delta|+\upsilon>0.
\end{equation}
The parameter $\upsilon$ can be either $\upsilon>0$ or
$\upsilon<0$ and in the simplest case it is not function of the
energy gap $|\Delta|$ ($\upsilon=0$ is a trivial case
corresponding to BCS theory). Moreover it is supposed that
$\upsilon$ does not depend on temperature essentially like
parameters of electron-phonon interaction. The condition
$2|\Delta|+\upsilon>0$ ensures stability of the Cooper pairs
(bound state of the electrons is energetically favorable),
otherwise transformation (\ref{2.1}) has not sense and such SC
state can not exist. If $\upsilon<0$ then the breaking of a Cooper
pair lowers energy of the molecular structure (or creation of the
pair raises the energy). In this case the pairs become less
stable. If $\upsilon>0$ then the breaking of the pair raises the
energy (creation of the pair lowers the energy). In this case the
pairs become more stable. The Hamiltonian corresponding to the
transformations (\ref{2.1}) is
\begin{eqnarray}\label{2.1a}
    \widehat{H}&=&\widehat{H}_{\texttt{BCS}}+\widehat{H}_{\upsilon}=\sum_{\textbf{k},\sigma}\varepsilon(k)a_{\textbf{k},\sigma}^{+}a_{\textbf{k},\sigma}
    -\frac{\lambda}{V}\sum_{\textbf{k},\textbf{p}}a_{\textbf{p}\uparrow}^{+}a_{-\textbf{p}\downarrow}^{+}a_{-\textbf{k}\downarrow}a_{\textbf{k}\uparrow}
    -\frac{\upsilon}{2}\sum_{\textbf{k}}\left[\frac{\Delta}{|\Delta|}a_{\textbf{k}\uparrow}^{+}a_{-\textbf{k}\downarrow}^{+}
    +\frac{\Delta^{+}}{|\Delta|}a_{-\textbf{k}\downarrow}a_{\textbf{k}\uparrow}\right],
\end{eqnarray}
where $\widehat{H}_{\texttt{BCS}}$ is BCS Hamiltonian: kinetic
energy + pairing interaction ($\lambda>0$). The combinations
$a_{\textbf{k}\uparrow}^{+}a_{-\textbf{k}\downarrow}^{+}$ and
$a_{-\textbf{k}\downarrow}a_{\textbf{k}\uparrow}$ are creation and
annihilation of Cooper pairs operators. The field $\upsilon$ is
called as the \textit{external pair potential}, since the
potential is imposed on the electron subsystem by the structural
elements of matter. The potential \emph{essentially renormalizes
the order parameter} $\Delta$ which is determined from
self-consistency equation:
\begin{equation}\label{2.2}
\Delta=g\int_{-\omega}^{\omega}d\xi
    \frac{\Delta\left(1+\frac{\upsilon}{2|\Delta|}\right)}
    {2\sqrt{\xi^{2}+|\Delta|^{2}\left(1+\frac{\upsilon}{2|\Delta|}\right)^{2}}}
    \tanh\left(\frac{\sqrt{\xi^{2}+|\Delta|^{2}\left(1+\frac{\upsilon}{2|\Delta|}\right)^{2}}}{2T}\right).
\end{equation}
We can see that only the electron-electron coupling is the cause
of superconductivity (if $g\equiv\lambda\nu_{F}=0$ then
$\Delta=0$), but not the potential $\upsilon$. If $\upsilon=0$ we
have the usual self-consistency equation in BCS theory. If
$\upsilon>0$ then at large temperatures $T\gg T_{c}(\upsilon=0)$
the energy gap tends to zero asymptotically as $1/T$ with
increasing of temperature:
\begin{equation}\label{2.3}
    |\Delta|=\frac{g\omega\upsilon}{4T}.
\end{equation}
Thus, formally, the critical temperature is equal to infinity,
however the energy gap remains finite quantity - Fig.\ref{Fig6}.

\begin{figure}[h]
\includegraphics[width=16cm]{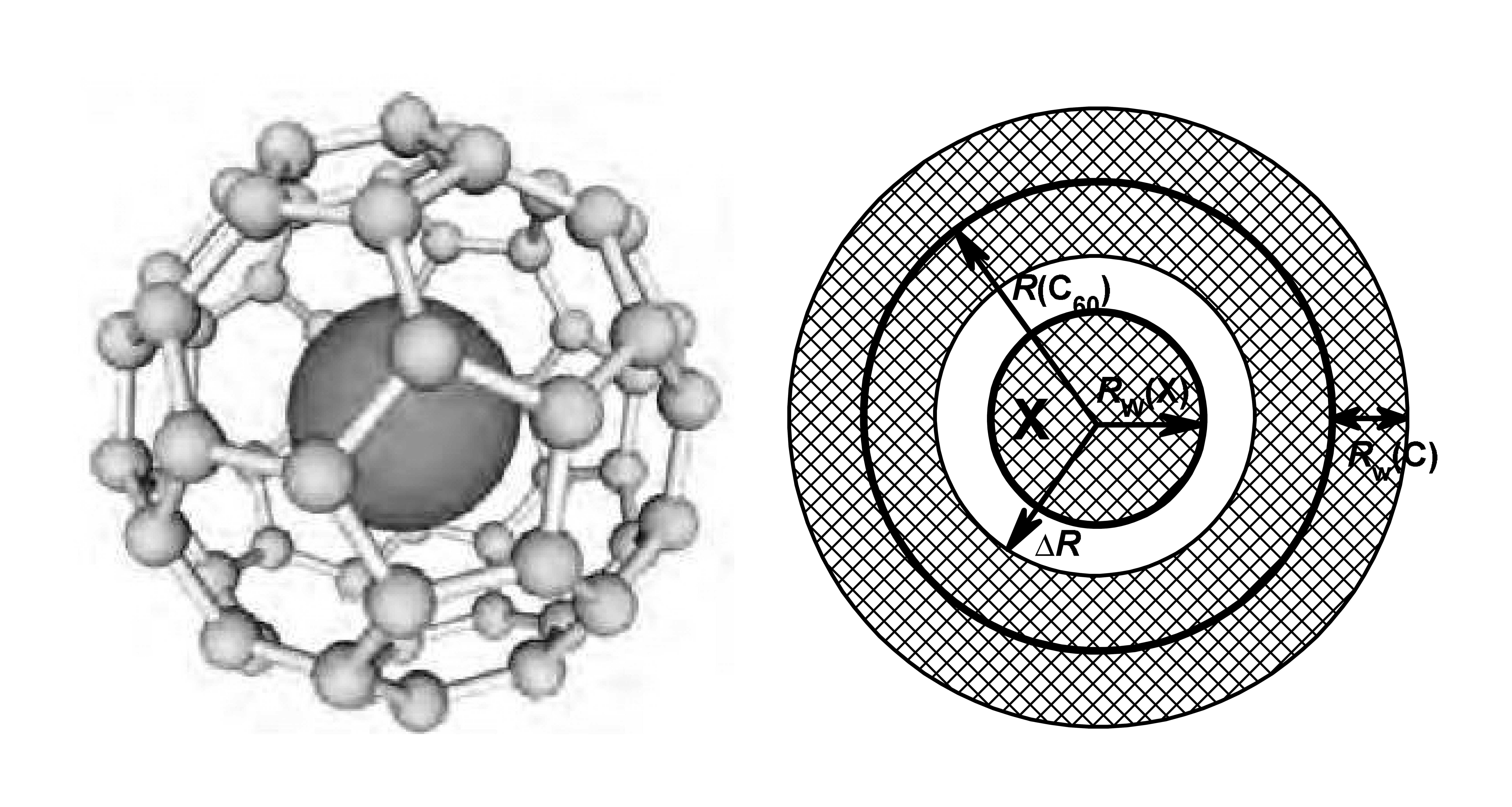}
\caption{An endohedral fullerene $\texttt{X@C}_{60}$ and
cross-section of the endohedral fullerene. The carbon cage can be
considered as a spherical layer of thickness
$2R_{\texttt{W}}(\texttt{C})$ and central radius
$R(\texttt{C}_{60})$. The central atom $\texttt{X}$ placed into
the inner cavity of radius $\Delta R$ is a noble gas atom of Van
der Waals radius $R_{\texttt{W}}(\texttt{X})$.} \label{Fig7}
\end{figure}
\begin{figure}[h]
\includegraphics[width=16cm]{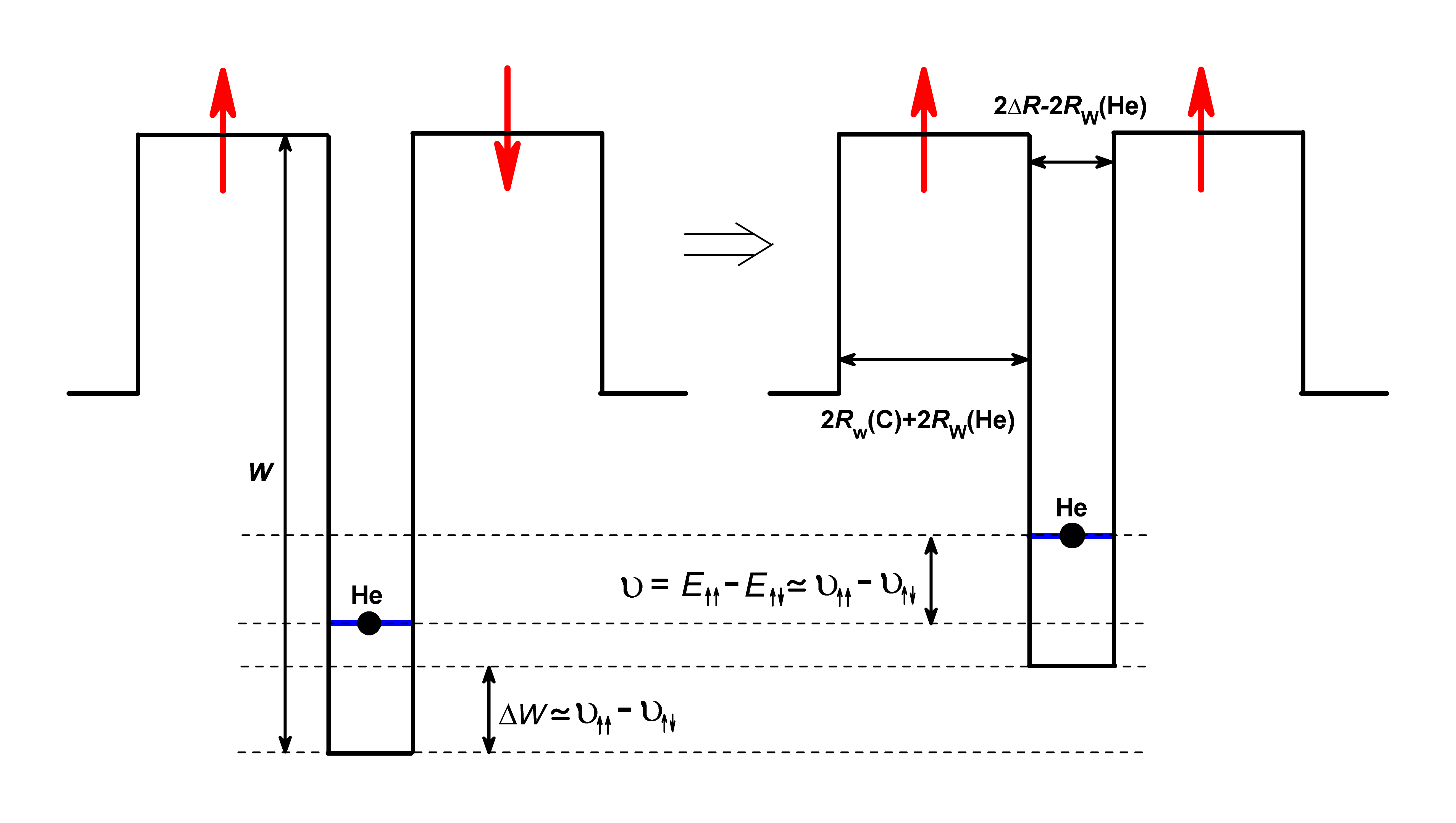}
\caption{Process of breaking of an electron pair localized on a
endohedral fullerene $\texttt{He@C}_{60}$. The process is
accompanied by change of energy of a central helium atom
$E_{\uparrow\uparrow}-E_{\uparrow\downarrow}\simeq\upsilon_{\uparrow\uparrow}-\upsilon_{\uparrow\downarrow}$
(if $\Delta W\ll W$).} \label{Fig8}
\end{figure}

In this work based on alkali-doped fulleride we propose
realization of this model. As it has been show above, in this
material a Cooper pair is formed on a molecule of size
$R=3.55\texttt{A}$ as result a of the el.-vib. interaction and
suppression of hopping between molecules by one-cite Coulomb
interaction $U$. This situation is fundamentally different from
superconductivity in metals, where the size of a Cooper pair is
macroscopic quantity $\sim 10^{3}\ldots 10^{4}\texttt{A}$. Let us
consider some features of the molecular structure
$\texttt{C}_{60}$. The Van der Waals radius of a carbon atom is
$R_{\texttt{W}}(\texttt{C})=1.70\texttt{A}$. Thus a fullerene has
an inner cavity in its center of size $\Delta
R=R(\texttt{C}_{60})-R_{\texttt{W}}(\texttt{C})=1.85\texttt{A}$. A
noble gas atom $\texttt{X}$ can be trapped in a carbon cage in the
inner cavity (Fig.\ref{Fig7}) - we have an endohedral complex
$\texttt{X@C}_{60}$
\cite{breton,cios,saund,rubin,dennis,jime,patchk}. Since for a
helium atom $R_{\texttt{W}}(\texttt{He})=1.40\texttt{A}<\Delta R$
it's electronic shell does not make hybridized orbitals with
electronic shells of the carbon cage. If helium atoms are placed
into each fullerene molecule in alkali-doped fulleride, then we
have a hypothetical material $\texttt{A}_{3}\texttt{He@C}_{60}$.
Electronic properties of $\texttt{A}_{3}\texttt{He@C}_{60}$ must
be identical to electronic properties of
$\texttt{A}_{3}\texttt{C}_{60}$. Changes in oscillation spectrum
of a fullerene can be neglected.

As noted above, in the endohedral fullerene the helium atom is
trapped in a carbon cage and it interacts with the cage by Van der
Waals force because $R_{\texttt{W}}(\texttt{He})<\Delta R$. The
carbon cage is barrier of width
$2R_{\texttt{W}}(\texttt{C})+2R_{\texttt{W}}(\texttt{He})$ for the
helium atom and there is the well of width $2\Delta
R-2R_{\texttt{W}}(\texttt{He})=0.9\texttt{A}$ in the center of
fullerence - Fig.\ref{Fig8}. Let depth of the well is $W\sim
70\texttt{meV}$ \cite{breton} and oscillation frequency (the
average distance between energy levels) of the trapped helium atom
is $\Omega\sim 60\ldots 120\texttt{cm}$. Van der Waals interaction
depends on electron configuration of interacting subsystems
(polarizability depends on the electron configurations). In
alkali-doped fullerenes the alkali metal atoms give valent
electrons to fullerene molecules. Then the interaction of the
trapped helium atom with the carbon cage can depend on a state of
the excess electrons on the surface of a molecule
$\texttt{C}_{60}$. Any two excess electrons can be in the paired
state (\ref{1.4a}) or in the normal state (\ref{1.4b}). Let the
interaction energy for the paired state is
$\upsilon_{\uparrow\downarrow}$ and the energy for the normal
state is $\upsilon_{\uparrow\uparrow}$. If
$\upsilon_{\uparrow\downarrow}<\upsilon_{\uparrow\uparrow}$ then a
molecule $\texttt{X}@\texttt{C}_{60}^{-n}$ has lower energy if the
excess electrons are in the paired state than the energy if the
electrons are in the normal state. If
$\upsilon_{\uparrow\downarrow},\upsilon_{\uparrow\uparrow}\ll W$
then change of depth of the well is $\Delta
W\simeq\upsilon_{\uparrow\uparrow}-\upsilon_{\uparrow\downarrow}$
at breaking of the pair. If
$\upsilon_{\uparrow\downarrow},\upsilon_{\uparrow\uparrow}\ll
\Omega$ then the change of the helium atom's energy is
$E_{\uparrow\uparrow}-E_{\uparrow\downarrow}\simeq\upsilon_{\uparrow\uparrow}-\upsilon_{\uparrow\downarrow}$.
Hence a function
\begin{equation}\label{2.4}
    \upsilon=\upsilon_{\uparrow\uparrow}-\upsilon_{\uparrow\downarrow}
\end{equation}
plays role of the external pair potential.
The
Hamiltonian corresponding to such system is
\begin{eqnarray}\label{2.4a}
\widehat{H}_{eff}&=&\sum_{\langle
ij\rangle}\sum_{m}\sum_{\sigma}\left(t_{ij}^{mm}+(\varepsilon_{m}-\mu)\right)
a^{+}_{im\sigma}a_{jm\sigma}\nonumber\\
&+&\left(U-U'+3K-U_{vib}\right)\sum_{i}\sum_{m}n_{im\uparrow}n_{im\downarrow}
-U_{vib}'\sum_{i}\sum_{m'\neq
m}a_{im\uparrow}^{+}a_{im'\uparrow}a_{im\downarrow}^{+}a_{im'\downarrow}\nonumber\\
&-&\frac{\upsilon}{2}\sum_{i}\sum_{m}\left[\frac{\Delta}{|\Delta|}a_{im\uparrow}^{+}a_{im\downarrow}^{+}
+\frac{\Delta^{+}}{|\Delta|}a_{im\downarrow}a_{im\uparrow}\right],
\end{eqnarray}
which can be reduced to the Hamiltonian (\ref{2.1a}) (where
$\lambda=3U_{vib}-U+U'-3K$) in reciprocal (momentum) space with
help of transitions (\ref{1.8}) and the assumption
$U_{vib}'=U_{vib}$.

The Van der Waals interaction is interaction due to virtual
transitions of the Cooper pair's electrons from a triply
degenerate level $t_{1u}$ ($l=5$) to levels $t_{1g}$ ($l=5$),
$h_{g},t_{2u},h_{u}$ ($l=6$), $g_{g},g_{u},t_{g}$ ($l=7$),
$\ldots$, where $l$ is an orbital index for $\pi$-electrons (see
Fig.\ref{Fig1}):
\begin{equation}\label{2.5}
    \Phi_{0\uparrow\downarrow}\equiv
    \Omega_{l=5,\gamma}(\textbf{R}_{1})\Omega_{l=5,\gamma}(\textbf{R}_{2})
    \longleftrightarrow
    \frac{1}{\sqrt{2}}\left[\Omega_{l,\gamma'}(\textbf{R}_{1})\Omega_{l=5,\gamma}(\textbf{R}_{2})
    +\Omega_{l=5,\gamma}(\textbf{R}_{1})\Omega_{l,\gamma'}(\textbf{R}_{2})\right]\equiv\Phi_{k\uparrow\downarrow}
    \footnotemark\footnotetext{we should write the ground state wave function as
    $\frac{1}{\sqrt{3}}\sum_{\gamma=1}^{3}\Omega_{l=5,\gamma}\Omega_{l=5,\gamma}$. Accordingly the excited state wave function
    $\Phi_{k\uparrow\downarrow}$ will have more cumbersome form. We suppose that results of calculations do not depend on the quantum index
    $\gamma$ (that will be confirmed with the further calculations).
    Hence for convenience we can write the ground state wave functions in forms (\ref{2.5}) and (\ref{2.9}) fixing the index $\gamma$.},
\end{equation}
and of the helium atom from a level $1s$ to levels
$2s,2p,3s,3p,3d,\ldots$:
\begin{eqnarray}\label{2.6}
    &&\Psi_{0}\equiv
    f_{1,0}(r_{1})Y_{0,0}(\textbf{r}_{1})f_{1,0}(r_{2})Y_{0,0}(\textbf{r}_{2})\nonumber\\
    &&\longleftrightarrow
    \frac{1}{\sqrt{2}}\left[f_{n,\widetilde{l}}(r_{1})Y_{\widetilde{l},\widetilde{m}}(\textbf{r}_{1})f_{1,0}(r_{2})Y_{0,0}(\textbf{r}_{2})
+f_{1,0}(r_{1})Y_{0,0}(\textbf{r}_{1})f_{n,\widetilde{l}}(r_{2})Y_{\widetilde{l},\widetilde{m}}(\textbf{r}_{2})\right]\equiv\Psi_{p},
\end{eqnarray}
where the index $\gamma$ labels the irreducible representation of
the icosahedral symmetry group; $n,l,m$ are principal, orbital and
magnetic quantum numbers accordingly; $f_{n,l}(r)$ is a radial
wave function, $Y_{l,m}$ is a spherical wave functions. $\Phi_{0}$
and $\Psi_{0}$ are ground-states of a Cooper pair and a helium
atom accordingly. $\Phi_{k}$ and $\Psi_{p}$ are the excited states
of the Cooper pair and the helium atom accordingly, $k$ and $p$
are sets of quantum indices of the corresponding exited states.
$\textbf{R}_{1}$ and $\textbf{R}_{2}$ are radius-vectors of
electrons of the Cooper pair, and
$|\textbf{R}_{1}|\approx|\textbf{R}_{2}|\approx R$ since the
Cooper pair is on surface of the molecule. $\textbf{r}_{1}$ and
$\textbf{r}_{2}$ are radius-vectors of electrons of the helium
atom, and $\langle r\rangle=0.31\texttt{A}\ll R=3.55\texttt{A}$ -
the atom is much less than the fullerene molecule. Signs "+" in
the sums are caused by the fact that the ground states of both the
Cooper pair and the helium atom are singlet, and transitions
between singlet and triplet states are allowed only due the
spin-orbit interaction, but this interaction can be neglected.
Energy of the Van der Waals interaction is defined with the second
order correction:
\begin{equation}\label{2.7}
    \upsilon_{\uparrow\downarrow}=\sum_{k,p}\frac{|\langle \Phi_{k\uparrow\downarrow},\Psi_{p}|\widehat{V}|\Psi_{0},\Phi_{0\uparrow\downarrow}\rangle|^{2}}
    {E_{0}+\widetilde{E}_{0}-E_{k}-\widetilde{E}_{p}},
\end{equation}
where the summation is done over the indexes of all possible
excited states of the Cooper pair $k$ and of the helium atom $p$;
$E_{0}$ and $\widetilde{E}_{0}$ are ground state energies of the
Cooper pair and the helium atom accordingly, $E_{k}$ and
$\widetilde{E}_{p}$ are energies of the corresponding excited
states. Since $E_{0}<E_{k},\widetilde{E}_{0}<\widetilde{E}_{p}$
then $\upsilon_{\uparrow\downarrow}<0$. An operator of the
interaction is
\begin{equation}\label{2.8}
   \widehat{V}(\textbf{R}_{1},\textbf{R}_{2},\textbf{r}_{1},\textbf{r}_{2})=
   \left(\frac{e^{2}}{|\textbf{R}_{1}-\textbf{r}_{1}|}+\frac{e^{2}}{|\textbf{R}_{1}-\textbf{r}_{2}|}+
   \frac{e^{2}}{|\textbf{R}_{2}-\textbf{r}_{1}|}+\frac{e^{2}}{|\textbf{R}_{2}-\textbf{r}_{2}|}
   -\frac{2e^{2}}{|\textbf{R}_{1}|}-\frac{2e^{2}}{|\textbf{R}_{2}|}\right)/\varepsilon_{\infty},
   \end{equation}
where $\varepsilon_{\infty}\approx 5\ldots 6$ \cite{degior} is the
high-frequency dielectric function because the Van der Waals
interaction is result of virtual transitions between atomic
(molecular) levels (between molecular levels of $\texttt{C}_{60}$
- $\sim 3\texttt{eV}$, between the levels of a helium atom - $\sim
20\texttt{eV}$), while the plasma frequency in
$\texttt{A}_{3}\texttt{C}_{60}$ is $\simeq 1.1\texttt{eV}$. This
is a very rough estimation of the screening, because the
dielectric function is a continual characteristic of a medium,
however we are dealing with intramolecular process. The problem of
interaction of the trapped atom with the fullerene cage occupied
additional electrons on LUMO in an alkali doped fulleride requires
a separate study.

To calculate the energy of Van der Waals interaction if electrons
are in the normal state (\ref{1.4b}) we can use an antisymmetric
wave function:
\begin{eqnarray}\label{2.9}
    \Phi_{0\uparrow\uparrow}\equiv
    \frac{1}{\sqrt{2}}\left[\Omega_{l=5,\gamma_{1}}(\textbf{R}_{1})\Omega_{l=5,\gamma_{2}}(\textbf{R}_{2})
    -\Omega_{l=5,\gamma_{1}}(\textbf{R}_{2})\Omega_{l=5,\gamma_{2}}(\textbf{R}_{1})\right]\nonumber\\
    \longleftrightarrow
    \frac{1}{\sqrt{2}}\left[\Omega_{l,\gamma'}(\textbf{R}_{1})\Omega_{l=5,\gamma_{2}}(\textbf{R}_{2})
    -\Omega_{l,\gamma'}(\textbf{R}_{2})\Omega_{l=5,\gamma_{2}}(\textbf{R}_{1})\right]\equiv\Phi_{k\uparrow\uparrow}.
\end{eqnarray}
Energy of the Van der Waals interaction is
\begin{equation}\label{2.10}
    \upsilon_{\uparrow\uparrow}=\sum_{k,p}\frac{|\langle
    \Phi_{k\uparrow\uparrow},\Psi_{p}|\widehat{V}|\Psi_{0},\Phi_{0\uparrow\uparrow}\rangle|^{2}}
    {E_{0}+\widetilde{E}_{0}-E_{k}-\widetilde{E}_{p}}.
\end{equation}
It should be noted that due to electroneutrality of a helium atom
and small size of the atom compared to radius of the molecule
$\langle r\rangle=0.31\texttt{A}\ll R=3.55\texttt{A}$ we have that
$\langle 00|V|00\rangle=0$ - the first order process can be
neglected. We neglect the higher order processes, i.e.
non-additivity of Van der Waals interaction. Moreover we can
neglect the exchange processes between the helium atom and
electrons on the molecule's surface, that cannot be done, for
example, for atoms $\texttt{Ar}$ and $\texttt{Xe}$ due to overlap
of the atom's wave functions with the wave functions of electrons
of the fullerene cage \cite{verkh,mad,pott,Ito1}.

Let us compare the energies $\upsilon_{\uparrow\downarrow}$ and
$\upsilon_{\uparrow\uparrow}$. To do this it is necessary to
notice that the ground states of electrons on surface of a
fullerene molecule have an important property. As well known
\begin{equation}\label{2.11}
    \left\langle\Omega_{l,\gamma}|\Omega_{l',\gamma'}\right\rangle=\delta_{ll'}\delta_{\gamma\gamma'},
    \quad\left\langle
    f_{n,l}Y_{l,m}|f_{n',l'}Y_{l',m'}\right\rangle=\delta_{nn'}\delta_{ll'}\delta_{mm'}.
\end{equation}
Then from Eq.(\ref{2.9}) we can see that the high-spin ground
state $\Phi_{0\uparrow\uparrow}$ (normal state) is superposition
of mutually orthogonal states with the weights $1/\sqrt{2}$ each,
unlike the low-spin ground state (paired state)
$\Phi_{0\uparrow\downarrow}$ - Eq.(\ref{2.5}). We have the
following matrix elements:
\begin{equation}\label{2.12}
    \langle \Phi_{k\uparrow\downarrow},\Psi_{p}|\widehat{V}|\Psi_{0},\Phi_{0\uparrow\downarrow}\rangle=2e^{2}\int d\textbf{R}_{1}d\textbf{r}_{1}
    \frac{1}{|\textbf{R}_{1}-\textbf{r}_{1}|}\Omega_{l,\gamma'}^{+}(\textbf{R}_{1})\Omega_{5,\gamma}^{+}(\textbf{R}_{1})
    f_{n,\widetilde{l}}(r_{1})Y_{\widetilde{l},\widetilde{m}}^{+}(\textbf{r}_{1})f_{1,0}(r_{1})Y_{0,0}(\textbf{r}_{1})/\varepsilon_{\infty},
\end{equation}
and
\begin{equation}\label{2.13}
    \langle \Phi_{k\uparrow\uparrow},\Psi_{p}|\widehat{V}|\Psi_{0},\Phi_{0\uparrow\uparrow}\rangle=\sqrt{2}e^{2}\int d\textbf{R}_{1}d\textbf{r}_{1}
    \frac{1}{|\textbf{R}_{1}-\textbf{r}_{1}|}\Omega_{l,\gamma'_{1}}^{+}(\textbf{R}_{1})\Omega_{5,\gamma_{1}}^{+}(\textbf{R}_{1})
    f_{n,\widetilde{l}}(r_{1})Y_{\widetilde{l},\widetilde{m}}^{+}(\textbf{r}_{1})f_{1,0}(r_{1})Y_{0,0}(\textbf{r}_{1})/\varepsilon_{\infty},
\end{equation}
where $l\geq5$ (if $l=5$ then $\gamma'$ and $\gamma_{1}'$
correspond to states in $t_{2u}$ level - Fig.\ref{Fig1}), $n>1,
\widetilde{l}>0$, $d\textbf{R}\equiv d\varphi\sin\theta
d\theta,d\textbf{r}\equiv r^{2}dr d\varphi\sin\theta d\theta$. As
noted above we have supposed that results of calculations do not
depend on the quantum indexes $\gamma,\gamma_{1}$ in the ground
state wave functions. Hence we can fix $\gamma$ by one of the
indexes from $t_{1u}$. Let us determine the corresponding matrix
elements assuming $\gamma_{1}=\gamma$ and $\gamma'_{1}=\gamma'$
(the indexes $l,\gamma',\gamma'_{1}$ run through all excited
states by which the summation is done in
Eqs.(\ref{2.7},\ref{2.10})). Then we have
\begin{equation}\label{2.14}
    \langle \Phi_{k\uparrow\downarrow},\Psi_{p}|\widehat{V}|\Psi_{0},\Phi_{0\uparrow\downarrow}\rangle
    =\sqrt{2}\langle \Phi_{k\uparrow\uparrow},\Psi_{p}|\widehat{V}|\Psi_{0},\Phi_{0\uparrow\uparrow}\rangle
\end{equation}
for the corresponding matrix elements. Hence we obtain
\begin{equation}\label{2.15}
    \upsilon_{\uparrow\downarrow}=2\upsilon_{\uparrow\uparrow} \footnotemark\footnotetext{we can select another
$\gamma_{1}\neq\gamma$ but then we cannot determine the
corresponding matrix elements with the relation (\ref{2.14})
because $\int d\textbf{R}_{1}
    \frac{1}{|\textbf{R}_{1}-\textbf{r}_{1}|}\Omega_{l,\gamma'}^{+}(\textbf{R}_{1})\Omega_{5,\gamma}^{+}(\textbf{R}_{1})\neq \int d\textbf{R}_{1}
    \frac{1}{|\textbf{R}_{1}-\textbf{r}_{1}|}\Omega_{l,\gamma'_{1}}^{+}(\textbf{R}_{1})\Omega_{5,\gamma_{1}}^{+}(\textbf{R}_{1})$
    if $\gamma_{1}\neq\gamma,\gamma'_{1}\neq \gamma_{1}$. However the relation $\upsilon_{\uparrow\downarrow}\approx2\upsilon_{\uparrow\uparrow}$
     can be checked by direct calculation looking over all possible transitions.}.
\end{equation}
Since the sign of Van der Waals interactions
(\ref{2.7},\ref{2.10}) is always negative
$\upsilon_{\uparrow\downarrow}<0,\upsilon_{\uparrow\uparrow}<0$,
then
\begin{equation}\label{2.16}
    \upsilon=-\upsilon_{\uparrow\uparrow}=-\frac{1}{2}\upsilon_{\uparrow\downarrow}>0.
\end{equation}
Thus an endohedral fullerene molecule
$\texttt{He}@\texttt{C}_{60}$ with excess electrons has a lower
energy if the electrons are in the paired state (\ref{1.4a}) than
energy if the electrons are in the state according to Hund's rule
(\ref{1.4b}). This means that the potential $\upsilon$
renormalizes the order parameter $\Delta$ so that the asymptotic
(\ref{2.3}) occurs.

To estimate $\upsilon$ a fullerene molecule can be considered as a
sphere, that simplifies the calculation of the matrix elements
(\ref{2.12},\ref{2.13}). Then the wave functions on the
fullerene's surface $\Omega_{l,\gamma}$ are spherical wave
functions $Y_{l,m}$. Ground sate of an electron on the molecule is
state with $l=5$, which is degenerated in $m=-l\ldots l$ . The
electron make virtual transitions to state with $l=6,7,\ldots$.
Energy of each level is
\begin{equation}\label{2.17}
    E_{l}=\frac{\hbar^{2}l(l+1)}{2m_{e}R^{2}},
\end{equation}
that gives  $E_{5,m}-E_{6,m'}\approx 3.6\texttt{eV}$. This value
is order of distance between nearest molecular orbitals in a
fullerene. For helium atom we have
$\widetilde{E}_{1,0,0}-\widetilde{E}_{n,\widetilde{l},\widetilde{m}}\approx
21\texttt{eV}$. Calculation shows that $\upsilon$ is almost
independent of the quantum index $m$:
\begin{equation}\label{2.18}
    \upsilon_{\uparrow\downarrow}(m=0\ldots\pm 5)\approx -80/\varepsilon_{\infty}^{2}\texttt{K},
\end{equation}
hence $\upsilon=-40/\varepsilon_{\infty}^{2}\texttt{K}$. For
example, at $T=300\texttt{K}$ with help Eq.(\ref{2.3}) we have
$\Delta\simeq0.5\texttt{K}$. However, as noted above the
description of the screening with the dielectric function
$\varepsilon_{\infty}$ is a very rough estimation, because this
function is a continual characteristic of a medium, however we are
dealing with intramolecular process.

In connection with the obtained results it should be noted that in
works \cite{Ito2,Ito3} it had been reported about the synthesis of
the first endohedral fullerene superconductors
$\texttt{A}_{3}\texttt{Ar}@\texttt{C}_{60}$ having critical
temperatures on $2-3$ kelvins less than critical temperatures of
the pure materials $\texttt{A}_{3}\texttt{C}_{60}$. The Van der
Waals radius of $\texttt{Ar}$ is $R_{\texttt{W}}=1.88\texttt{A}$
that is slightly more than radius $\Delta R=1.85\texttt{A}$ of the
inner cavity in center of a fullerene molecule. In this case an
overlap of the argon atom's wave functions with the wave functions
of electrons of the fullerene cage occurs, hence the exchange
interaction plays a role (the electron shell of a noble gas atom
is embedded in the electron shell of fullerene)
\cite{verkh,mad,pott,Ito1}. The radii of $\texttt{Kr}$ and
$\texttt{Xe}$ are much larger, hence role of the exchange
interaction is more significant. Thus the influence of
$\texttt{Ar},\texttt{Kr},\texttt{Xe}$ requires special
consideration that goes beyond the present work. On the other
hand, it should be noted that the Van der Waals radius of hydrogen
molecule is $R_{\texttt{W}}(\texttt{H}_{2})\approx
R_{\texttt{W}}(\texttt{H})+R_{\texttt{\texttt{cov}}}(\texttt{H})=1.57\texttt{A}<\Delta
R$, where $R_{\texttt{\texttt{cov}}}$ is a covalent radius. Thus
the hydrogen molecule can be placed in the inner cavity of a
fullerene without overlap of the electron shells like a helium
atom. Hence the system
$\texttt{A}_{3}\texttt{H}_{2}@\texttt{C}_{60}$ can be similar to
the system $\texttt{A}_{3}\texttt{He}@\texttt{C}_{60}$.

\section{Summary}\label{concl}

We have reviewed conductivity and superconductivity of
alkali-doped fullerides $\texttt{A}_{n}\texttt{C}_{60}$
($\texttt{A}=\texttt{K},\texttt{Rb},\texttt{Cs}$, $n=1\ldots 5$)
while these materials would have to be antiferromagnetic Mott
insulators because the one-site Coulomb interaction is lager than
bandwidth $U\sim 1\texttt{eV}>W\sim 0.5\texttt{eV}$ and electrons
on a molecule have to be distributed over molecular orbitals
according to Hund's rule (\ref{1.4b}). We have found important
role of the triply degeneracy of LUMO ($t_{1u}$ level), small
hopping between neighboring molecules and the coupling of
electrons to Jahn-Teller modes. The coupling to Jahn-Teller modes
$U_{vib}$ cannot compete with the the Coulomb repulsion $U\gg
U_{vib}$, but it can compete with the Hund coupling
$U-U'+3K\approx5K\sim U_{vib}$ (where the exchange energy is much
less than direct Coulomb interaction $K\ll U$) due to the
threefold degeneration of the energy level $t_{1u}$. This allows
to form the local pairs (a pair is formed on a molecule -
Eq.(\ref{1.4a})). The formation of local pairs and the coupling to
Jahn–Teller modes radically change conductivity of these
materials: they can make $\frac{\Delta E}{W}<1$, where $\Delta E$
is energy change at transfer of an electron from one site to
neighboring site, while without accounting of the electron-vibron
interaction we have $\frac{\Delta E}{W}> 1$ that corresponds to
Mott insulator. We have shown that
$\texttt{A}_{1}\texttt{C}_{60},\texttt{A}_{5}\texttt{C}_{60}$ are
conductors but not superconductors.
$\texttt{A}_{3}\texttt{C}_{60}$ is superconductor.
$\texttt{A}_{2}\texttt{C}_{60},\texttt{A}_{4}\texttt{C}_{60}$ are
Mott insulators but their fulerenes do not have spin. This
mechanism can ensure conductivity of
$\texttt{A}_{3}\texttt{C}_{60}$ if bandwidth is $W\gtrsim
0.38\texttt{eV}$, that is
$\texttt{K}_{3}\texttt{C}_{60},\texttt{Rb}_{3}\texttt{C}_{60}$ are
conductors (and superconductors) but
$\texttt{Cs}_{3}\texttt{C}_{60}$ is Mott insulator at normal
pressure. We have shown that the system with the local pairing can
be effectively described by BCS theory for multi-band
superconductors. However in such system we have the effective
coupling constant as $g-\mu>0$ (where $g$ is determined with the
el.-vib. interaction and $\mu$ is determined with the Hund
coupling - Eq.(\ref{1.16})) unlike usual metal superconductors
where $g-\mu^{*}>0$ only (where $\mu^{*}$ is a Coulomb
pseudopotential with Tolmachev's reduction).

Based on the alkali-doped fulleride we have proposed realization
of the model of superconductivity with the external pair potential
formulated in \cite{grig}. In this model the potential essentially
renormalizes the order parameter so that if the pairing lowers the
energy of the molecular structure, then the energy gap tends to
zero asymptotically as $1/T$ with increasing of temperature -
Eq.(\ref{2.3}). Thus, formally, in this model the critical
temperature is equal to infinity, however the energy gap remains
finite quantity. For practical realization of this model we
propose a hypothetical superconductor on the basis of alkali-doped
fullerides using endohedral structures
$\texttt{He}@\texttt{C}_{60}$, where a helium atom is in the
center of each fullerene molecule. In an endohedral fullerene the
helium atom interacts with a carbon cage by Van der Waals force.
The interaction depends on a state of excess electrons on surface
of the molecule. We have shown that energy of the molecule if the
excess electrons on its surface are in the paired state
(\ref{1.4a})(when two electrons are in a state with the same
quantum numbers) is lower than the energy if the electrons are in
the normal state (\ref{1.4b})(when the electrons are in a state
with different quantum numbers and maximal spin). Thus difference
of the energies of the molecules plays a role of
the external pair potential. 


\end{document}